\documentclass[11pt]{article}
\usepackage{mydef2col}
\usepackage{kantlipsum,widetext}
\usepackage[utf8]{inputenc}
\usepackage[T1]{fontenc}
\usepackage[autostyle=true]{csquotes}
\usepackage{regexpatch}
\usepackage{cprotect}
\usepackage{soul}

\usepackage{tikz}
\usetikzlibrary{calc,decorations.markings}
\usetikzlibrary{arrows,patterns,positioning}

\usepackage{algorithm}
\usepackage{algpseudocode}
\newlength{\commentlen}
\newcommand{\CommentF}[1]{\Comment{\makebox[\commentlen][l]{\footnotesize{\textsc{#1}}}}}

\setlist{
  listparindent=\parindent,
  parsep=0pt,
}

\makeatletter
\DeclareRobustCommand{\pdot}{\mathbin{\mathpalette\pdot@\relax}}
\newcommand{\pdot@}[2]{%
  \ooalign{%
    $\m@th#1\circ$\cr
    \hidewidth$\m@th#1\cdot$\hidewidth\cr
  }%
}
\makeatother

\usepackage[colorinlistoftodos,prependcaption,textsize=tiny]{todonotes}
\makeatletter
\xpatchcmd{\@todo}{\setkeys{todonotes}{#1}}{\setkeys{todonotes}{inline,#1}}{}{}
\makeatother

\allowdisplaybreaks[4]
\graphicspath{{./Graphs/}}
\usepackage[backend=bibtex,style=authoryear,natbib=true]{biblatex}
\makeatletter
\newrobustcmd*{\parentexttrack}[1]{%
  \begingroup
  \blx@blxinit
  \blx@setsfcodes
  \blx@bibopenparen#1\blx@bibcloseparen
  \endgroup}

\AtEveryCite{%
  \let\bibcloseparen=\bibclosebracket}

\makeatother

\DeclareFieldFormat[article]{volume}{\textbf{#1}\addspace}
\renewbibmacro{volume+number+eid}{%
    \printfield{volume}%
    \setunit{\addcomma\space}%
    \printfield{number}%
    \printfield{eid}}

\newcommand{\barr}{\bar{r}}
\newcommand{\barR}{\bar{R}}

\newcommand{\Ind}{\mathbf{1}}

\def\calD{{\cal D}}
\def\calI{{\cal I}}
\def\calL{{\cal L}}
\def\calA{{\cal A}}

\def\tins{t_{inc}}

\def\Jnu{{J_{\nu}}}
\def\JnuM{{J_{|\nu|}}}
\def\Jnum{{J_{-\nu}}}
\def\Ynu{{Y_{|\nu|}}}

\def\lamIn{\lambda_{\mathrm{inh}}}
\def\blamIn{\bar{\lambda}_{\mathrm{inh}}}

 \title{Semi-analytical pricing of options written on SOFR futures}
\shorttitle{Semi-analytical pricing of options written on SOFR futures}
\author{
\authorstyle{Andrey Itkin\textsuperscript{1}
and Yerkin Kitapbayev\textsuperscript{2}
}
\newline\newline
\textsuperscript{1}
\institution{Tandon School of Engineering, New York University, USA,} \\
\textsuperscript{2}
\institution{Department of mathematics, Khalifa University of Science and Technology, Abu Dhabi, UAE.}
}

\date{\today}

\begin{document}

\maketitle

\lettrineabstract{In this paper, we propose a semi-analytical approach to pricing options on SOFR futures where the underlying SOFR follows a time-dependent CEV model. By definition, these options change their type at the beginning of the reference period: before this time, this is an American option written on a SOFR forward price as an underlying, and after this point, this is an arithmetic Asian option with an American style exercise written on the daily SOFR rates. We develop a new version of the GIT method and solve both problems semi-analytically, obtaining the option price, the exercise boundary, and the option Greeks. This work is intended to address the concern that the transfer from LIBOR to SOFR has resulted in a situation in which the options of the key money market (i.e., futures on the reference rate) are options without any pricing model available. Therefore, the trading in options on 3M SOFR futures currently ends before their reference quarter starts, to eliminate the final metamorphosis into exotic options.
}

\vspace{0.5in}
\section*{Introduction}

Traditionally, options, being derivative products, are used as hedging instruments. For instance, Fixed Income markets widely utilize options on LIBOR futures for hedging caps and floors. However, the migration from the London Interbank Offered Rate (LIBOR) to the Secured Overnight Financing Rate (SOFR) brought several problems to this process.

Historically,  SOFR futures and options on them have been introduced as the natural next step in the development of the SOFR ecosystem and launched in CME on January 6, 2020. Contract specifications for options written on SOFR futures can be found in \cite{SOFRRisk}, see also \cite{SOFR3m} as an introductory paper. SOFR Options can be executed on three venues: open outcry, CME Globex, and as a block trade submitted via CME ClearPort.  Each of these platforms offers customers access to deep and diverse pools of liquidity.

To better understand the peculiarities of the SOFR futures options, let us consider the specifications of this contract in more detail. As mentioned in \cite{HugginsSchaller}, CME Term SOFR rates are forward-looking interest rate estimates of overnight SOFR for reference periods starting $T+2$ from the date of publication, calculated and published for 1-month, 3-months, 6-months, and (since September 21, 2021) 12-months tenors. SOFR futures combine the daily rates before their reference period into a (forward) term rate. Hence, when dealing with SOFR futures, the situation is similar to LIBOR: SOFR futures transform the daily rates before their reference period into a term rate like LIBOR. And this is the reason behind the relatively easy transfer of concepts to the new reference rate. Via SOFR futures, one does not need to deal directly with daily SOFR, but with the daily SOFR aggregated by the future. But as soon as the reference period starts, one needs to deal with daily SOFR values.

\begin{figure}[!ht]
\hspace{-0.5in}
\captionsetup{width=0.8\linewidth}	
\begin{center}
\setlength{\fboxsep}{8pt}\fbox{
\begin{tikzpicture}
\def\axis{7.}
\def\refheight{0.5}
\def\textshift{-0.3}
\def\refnum{20}
\def\mult{1.1}
\def\step{2.*\mult*\axis/\refnum}
\def\refwidth{6.}
\def\refstart{-\mult*\axis + \refwidth*\step}
\def\refend{\refstart + 2*\refwidth*\step}

\draw (-\mult*\axis, 0) -- (\mult*\axis - \textshift, 0);
\node at (\mult*\axis - \textshift,\textshift){$t$};
\node at (\refstart, -\refheight + \textshift){$SR_{\mathrm{start}}$};
\node at (\refend, -\refheight + \textshift){$SR_{\mathrm{end}}$};
\node at (\refstart, 0.) {$\bullet$};
\node at (\refend, 0.) {$\bullet$};

\foreach \a in {1,...,\refnum}{
    \node at (-\mult*\axis + \a*\step, 0.){$\pdot$};
}

\node at (\refstart + \refwidth*\step, \textshift){$\mbox{\small Reference period of future}$};
\node at (\refstart + \refwidth*\step, -\textshift) {$\bm{\mbox{\small Daily SOFR rates}}$};
\draw[black, thick, fill=green!50, opacity=0.4] (\refstart, -\refheight) rectangle ++(2*\refwidth*\step, 2*\refheight);
\node at (-\mult*\axis + \step, \textshift){$S_{\mathrm{ins}}$};
\node at (-\mult*\axis + \step, 0){$\bullet$};
\node at (-\mult*\axis + 3*\step, \textshift){$S_{i}$};
\draw [cyan!30,fill=cyan!30] (-\mult*\axis + 3*\step, -3*\textshift) ellipse (3 and -\textshift);
\node at (-\mult*\axis + 3*\step, -3*\textshift) {$\bm{\mbox{\small Aggregated (forward) rate}}$};
\draw[->] (-\mult*\axis + 3*\step, 0) edge (-\mult*\axis + 3*\step, -2.*\textshift);

\end{tikzpicture}
}
\end{center}
\caption{Aggregation of the SOFR futures prices underlying SOFR futures options.}
\label{sofr_fut_agg}
\end{figure}
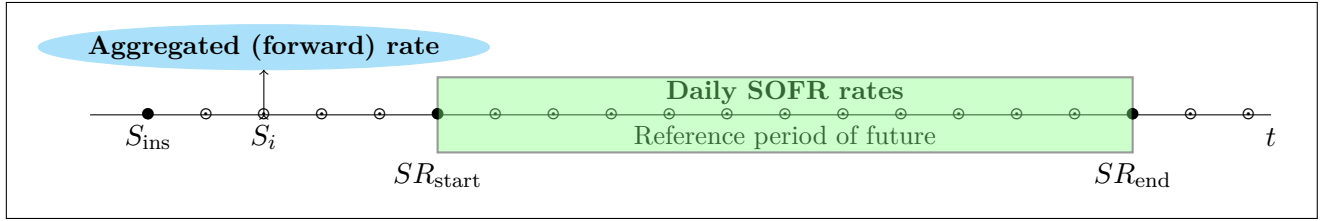

This is schematically illustrated in Fig.~\ref{sofr_fut_agg} (see also \cite{HugginsSchaller}).
Here $t$ is the time, $S_{\mathrm{inc}}$ is the inception date of the future contract, $S_i$ is some day before the reference period of the future, which starts at $SR_{\mathrm{start}}$ date and ends at the $SR_{\mathrm{end}}$ date. For the dates in the interval $[ S_{\mathrm{ins}}, SR_{\mathrm{start}} ]$ the SOFR future price is computed by aggregating the daily prices into a forward term rate for each date (e.g., $S_i$) in this interval. However, in the interval $[ S_{\mathrm{start}}, SR_{\mathrm{end}} ]$ this is switched to daily SOFR rates.

As far as the SOFR futures options is concerned, the above definition creates a new and rather complicated situation. Before the reference period, options on SOFR futures have a forward term rate as underlying, as do options on Eurodollar contracts. But during the reference period, they depend on the path of daily SOFR values during the reference period. Thus, from one day to the next, at the start of the reference period, the same option on the same SOFR future, which used to behave like a standard American option on an Eurodollar futures contract, suddenly becomes an Asian option of the American exercise style, since its underlying is now computed as an arithmetic average price of the SOFR daily rates. One might say that as SOFR futures enter their reference period, their aggregate function ends, and the options on them therefore transmogrify from standard options on a forward term rate to Asian options.

As per \cite{HugginsSchaller}, this can be considered as an uncontrolled experiment: the huge market for options on the reference rate is transferred from standard to highly exotic, from being accompanied by well-known pricing models to the absence of any pricing model. And one unintended consequence of the difficulties to pricing and hedging options on SOFR futures is their lesser suitability for hedging caps and floors – which may explain why market participants seem hesitant to migrate from Eurodollar to SOFR futures options.

In \cite{HugginsSchaller} another concern arises that both the markets using Asian options and the literature dealing with them have so far focused on Asian options of the European type only. Hence, the transfer from LIBOR to SOFR has resulted in a situation in which the options of the key money market (i.e., futures on the reference rate) are options without any pricing model available. Since the trading in options on 3M SOFR futures currently ends before their reference quarter starts, they do not experience the final metamorphosis into exotic options. This allows the transfer of the realized and implied volatility analysis from the unsecured to the secured yield curve, while  the similarity of specifications between options on SOFR and on Eurodollar futures enables executing spread positions easily.

However, the decision to let options on SOFR futures expire before they encounter the problem of becoming Asian during the reference period comes with the high cost of no options on futures being available until futures settlement. This means that it becomes impossible to hedge a cap or floor with a series of options on SOFR futures in the same manner as one can do with options on Eurodollar futures. In other words, the problem of a missing pricing formula for the hedging instrument has been “solved” by the hedging instrument itself being missing. Moreover, in contrast to the options on 3M SOFR futures, which end trading before the reference quarter begins, options on 1M SOFR futures end trading together with the underlying contract at the end of the reference month. Hence, these options are subject to the sudden switch to an Asian option with American-style exercise. The authors of \cite{HugginsSchaller} conclude that, while there currently is no way to price this hedging instrument during the reference month, at least the hedging instrument exists.

Thus, pricing options on SOFR futures seems to be an important problem that hasn't drawn sufficient attention in the existing literature. To mention, there exists an extensive literature on pricing Asian options of the American style exercise that is not directly related to options on SOFR futures. However, practically all of them deal with the Geometric Brownian Motion (GBM) process with constant coefficients for the underlying process, see, e.g. \cite{Ding2023, Xu2023} and references therein, and various papers on optimal stopping problems for the maximum of the GBM process, e.g., \cite{Gapeev2006} and references therein. As far as options on SOFR futures is concerned, we can mention a recent paper \cite{Mingyang2021} where the pricing of these options has been done numerically. However, it is rather obvious that building an efficient numerical algorithm (rather than one based on a Monte Carlo approach) is not an easy deal. Indeed, the underlying model is expected to be at least time-inhomogeneous to capture the term-structure of the real trading rates. At the same time, pricing Asian options in such models, especially with arithmetic averaging of the spot price, can be done just numerically (see some literature survey in \cite{HugginsSchaller}). On top of that, the American style of the option payoff brings even more problems since the exercise boundary of the option is not known and has to be computed simultaneously with the option price.

As shown in detail in \cite{Andersen2023}, various numerical problems associated with the finite difference (FD) and Monte Carlo (MC) approaches for pricing American options cannot be solved uniformly and easily and require even more sophisticated and tricky algorithms and ideas. The author highlights some typical problems with the existing numerical methods and demonstrates that the construction of efficient numerical algorithms for pricing American options is still an actual problem. Also, he concurs with our opinion that solving an integral equation for the exercise boundary is much more efficient than solving a pricing PDE numerically from both an efficiency and accuracy point of view.

Therefore, the main idea of this paper is to attack this pricing problem by using a quite different approach. By that, we mean semi-analytical pricing of American options, which considers the problem only in the continuation domain. Semi-analytical pricing of American options has drawn some attention within the last few years, see, e.g., \cite{CarrItkin2020jd, Kitapbayev2021, ItkinMuravey2024jd, Itkin2024jd} and references therein. In general, by a semi-analytical approach, we mean the following. To price an American option written on some underlying, e.g., an American Put option on a stock, we first consider a certain stochastic differential equation (SDE) with time-dependent coefficients, which describes the dynamics of the underlying. A partial differential equation (PDE) then can be derived such that the American option price solves it, subject to some terminal and boundary conditions. In many cases, this pricing PDE by a series of transformations can be reduced either to the Heat or Bessel equation with a general source term and moving boundaries. The corresponding solutions in the continuation region (where it is not optimal to exercise the American option) can be obtained analytically by using the Generalized Integral Transform (GIT) technique, \cite{ItkinLiptonMuraveyBook}, together with an extended version of Duhamel's principle, \cite{ItkinMuravey2024jd}. The thus obtained solution depends on the explicit form of the exercise boundary, which for American options is not known in advance. However, for every model considered in the above-mentioned papers, we derived a non-linear integral Volterra equation for the exercise boundary and provided some examples of how it can be efficiently solved numerically. It is worth mentioning, that in the past a similar result was known only for the Black-Scholes model with constant coefficients.

This approach is a nice alternative to standard numerical methods developed to price American options, e.g., a variety of FD methods, see \cite{ItkinBook} and references therein. As described, the core is that instead of calculating American option prices directly, one can find an explicit location of the option exercise boundary. This approach was advocated, e.g., in \cite{Andersen2016} for the Black-Scholes model with constant coefficients. It was shown that $S_B(t)$ solves an integral (Volterra) equation that can be solved numerically. The proposed numerical scheme can be implemented straightforwardly, and it converges at a speed several orders of magnitude faster than the other (previously mentioned) approaches. As mentioned in \cite{Itkin2024jd}, this approach also delivers various benefits for industrial applications when massive computations of American option prices have to be efficiently organized.

The rest of the paper is organized as follows. In \cref{Asian} we consider an auxiliary problem of pricing Asian options with an American payoff written on SOFR futures, assuming that the underlying SOFR is simulated by the time-inhomogeneous CEV model, where all coefficients of the model are some deterministic functions of the time $t$. \cref{American} discusses a similar auxiliary problem of pricing American options written on SOFR forwards, where again we use the time-homogeneous CEV model for the SOFR. \cref{experiments} presents the results of our experiments and simulations done by using this model. The final section concludes.

To the best of our knowledge, all the results obtained in the paper are new and have never been discussed in the existing literature.

\section{Pricing Asian-American options on SOFR futures} \label{Asian}

For modeling the evolution of daily SOFRs, we need to pick a time-inhomogeneous model for the instantaneous interest rate. In doing so, there is always a challenge between the tractability of the model on the one hand, and its capacity to capture the observed market behavior of the rates, e.g., the intrinsic convexity adjustments, skew and smile. This problem has drawn some attention in the literature, e.g., see \cite{Mercurio2018, TurfusRomeroBermudez2023} and references therein. In particular, as applied to modeling the SOFRs to capture skew and smile effects, in \cite{TurfusRomeroBermudez2023} the authors introduce the reduced variable $y_t$ defined as a time-homogeneous Ornstein-Uhlenbeck (OU) process
\begin{equation} \label{CEV}
	d y_t=-\alpha(t) y_t d t+\sigma(t) d W_t,
\end{equation}
\noindent where $W_t$ is the standard Wiener process, and $t \geq 0$ is the time. This auxiliary variable is related to the instantaneous short rate as  $r_t = r\left(t, y_t\right)$ where
\begin{equation} \label{process}
	r(y, t) = \barr(t) + R^*(t) + \frac{\sinh [\gamma(t)\left(y+y^*(t)\right) ]}{\gamma(t)}.
\end{equation}
Here $\barr, y^*: \mathbb{R}^{+} \rightarrow \mathbb{R}$ are the instantaneous forward rate and a skewness function, respectively, and $\sigma, \alpha, \gamma, R^*$ : $\mathbb{R}^{+} \rightarrow \mathbb{R}^{+}$ are deterministic functions of the time representing the volatility, the mean reversion rate, the smile factor and the convexity adjustment factor, respectively. In \cite{TurfusRomeroBermudez2023} they all assumed to be piecewise continuous and bounded, while the first assumption can be relaxed. It is also assumed that $y_0=0$, with $t=0$ being the "as of" date for which the model is calibrated. The function $R^*(t)$ is determined by calibration to the forward curve but will tend to zero in the zero volatility limit. The no-arbitrage constraint to determine this function reads
\begin{equation} \label{noArb}
E\left[e^{-\int_0^t r_s d s}\right]=\calD(0, t)
\end{equation}
\noindent under the martingale measure for $0<t \leq T_m$, where $T_m$ is the longest maturity date for which the model is calibrated, and
\begin{equation}
\calD\left(t_1, t_2\right)= e^{-\int_{t_1}^{t_2} \barr(s) d s}
\end{equation}
\noindent is the $t_1$-forward price of the $t_2$-maturity zero coupon bond (ZCB). The explicit condition on $R^*(t)$ that satisfies \eqref{noArb} has been given in \cite{TurfusRomeroBermudez2023}.
It is clear that at small $|\gamma(t)| \ll 1$ the instantaneous interest rate $r_t$ is linear in $y_t$, i.e. it is given by the Hull-White model, while otherwise it is non-linear.

Despite the various advantages brought by this model, for our purposes it is not sufficiently tractable. Therefore, instead, we proceed with a different model, which is a time-dependent constant elasticity of variance (CEV) model for $r_t$\footnote{
An alternative and extended (but time-homogeneous) model to capture the regulated dynamics of interest rates has been proposed in \cite{skewCEV2023}.}. This model is a one-dimensional diffusion process that solves a stochastic differential equation (e.g., see \cite{LinetskyMendozza2010, CarrItkinMuravey2020} and references therein)
\begin{align} \label{CEV1}
r_t &= \barr(t) + R^*(t) + y_t, \\
d y_t &= -\alpha(t) y_t dt + \sigma(t) y_t^{\beta+1} dW_t. \nonumber
\end{align}
Here $\beta$ is the elasticity parameter such that $\beta \in [-1,1)$. In case $\beta = 0$ this model is the time-dependent Black-Scholes model, while for $\beta = -1$ this is the Bachelier, or time-dependent OU model. The elasticity parameter $\beta$ controls the steepness of the skew (the larger the $|\beta|$, the steeper the skew), while the scale parameter $\sigma(t)$ fixes the at-the-money volatility level. This ability to capture the skew has made the homogeneous CEV model popular in various options markets, \cite{LinetskyMendozza2010}. On top of that, a time-inhomogeneous version of this model allows better calibration to the term-structure of SOFR futures.

Let us assume that all parameters of the model are known either as a continuous and bounded functions of time $t \in [0,\infty)$, or as a discrete set of $N$ values for some moments $t_i, \ i=1,\ldots,N$. Below, we look at an auxiliary problem of pricing options on SOFR futures, with the start date $t_0$ to be at the beginning of the reference period. For this period the option's underlying is an arithmetic average $R_A$ of the SOFRs over a given period, which is
\begin{equation}
R_A = \left[\sum_{i=1}^{d_b} \frac{r_i \times n_i}{N}\right] \times \frac{N}{d_c},
\end{equation}
\noindent where $d_b$ is the number of business days in the interest rate period $[t_0, t]$,  $d_c$ is the number of calendar days in this interest rate period, $r_i$ is the interest rate on the $i$-th business day, $n_i$ is the number of calendar days for which $r_i$ is applied, and $N$ is the number of calendar days in the year.

The simple average of SOFRs can be approximated by its continuous version
\begin{align} \label{contCompound}
R_A &= \left[\sum_{i=1}^{d_b} \frac{r_i \times n_i}{N}\right] \cdot\frac{N}{d_c}
\approx
\begin{cases}
r(t, y), & t =t_0, \\
\frac{1}{t-t_0} \int_{t_0}^{t} r_u d u, & t > t_0,
\end{cases}
\\
r(t_0, y)  &= \barR^*(t_0) + y, \quad y = y_{t_0}, \quad \barR^*(t) = R^*(t) + \barr(t). \nonumber
\end{align}
By introducing a new variable $z_t$
\begin{equation} \label{z}
z(t) = \frac{1}{t-t_0} \int_{t_0}^{t} r_u du,
\end{equation}
\noindent the average rate $R_A$ can be re-written in the form
\begin{equation} \label{singul}
R_A = z_t, \qquad z_{t_0} = \barR^*(t_0) + y.
\end{equation}
The last equality, which serves as the initial condition for $z_t$, originates from a standard argument that resolves the singularity at $t=t_0$ in the definition of $z_t$ in \eqref{z}, \cite{ZvanForsythVetzal1997}.

The \eqref{contCompound} is similar to the continuously compounded forward rate. As mentioned in \cite{Mingyang2021}, an advantage of simple interest is that it can be easier calculated in practice, and most existing systems can accommodate it.  On the other hand, simple interest makes pricing much more complicated because of the extra convexity adjustment that requires a stochastic interest rate model.

\subsection{Pricing PDE for American-Asian options in the continuation region}

Let us consider the price $P\left(t, y, z\right)$ of a Put option written on the SOFR future contract with $t \in [t_0, T]$, and $T$ being the maturity of the option (for instance, for options on 1M SOFR futures it is 1M ahead of $t_0$). Similar to \cite{RogersShi1995}, we assume that the strike $K$ and maturity $T$ of the option $T$ are fixed. Also, assume that the averaging of $z_t$ starts from $t_0: 0 \le t_0 < t < T$. To remind, this Put option is of the Asian-American type. By a standard argument, \cite{ingersoll:87, ZvanForsythVetzal1997},  in the continuation region (where it is not optimal to exercise the American option, see \cite{ItkinMuravey2024jd}) $P\left(t, y, z\right)$ the option price solves the following two-dimensional PDE\footnote{Since the SOFR futures options are settled in cache, it is not 100\% whether they have to be discounted back. For the moment, we leave the discounting term in the below PDE, but this can be easily relaxed. }
\begin{equation} \label{PDE}
\fp{P}{t} - \alpha(t) y \fp{P}{y} + \frac{1}{2} \sigma^2(t) y^{2 \beta+2}\sop{P}{y} + \frac{\barR^*(t) + y - z}{t-t_0}  \fp{P}{z} - [\barR^*(t) + y] P = 0.
\end{equation}
It has to be solved subject to the terminal condition at option maturity $t=T$
\begin{equation} \label{tc0}
	P(T, y, z) = (K - z(T))^+ = 0.
\end{equation}
The last equality  holds because we consider only the continuation region for the Put option. For instance, in the one-dimensional case this is the domain where $z_t \in [z_B(t), \infty)$ with $z_B(t)$ being the exercise boundary of the American option, see Fig.~1 in \cite{ItkinMuravey2024jd} and the corresponding discussion there. For our problem the exercise boundary becomes two-dimensional, which we further denote as $z_B(t,y)$. Thus, the full problem is set at the domain $(t,y,z) \in \Omega: [t_0, \infty) \times [y_l(t), \infty) \times [z_B(t,y), \infty)$, where $y_l(t)$ is some left boundary on $y_t$. Accordingly, we set the boundary conditions
\begin{align} \label{bc0}
P(t, y, z_B(t, y)) &= K - z_B(t, y), \qquad \left. \fp{}{z} P(t, y, z) \right|_{z = z_B(t, y)} = - 1, \\
P(t, y \uparrow \infty, z ) &= 0. \nonumber
\end{align}
The first and second equalities represent the "continuous-fit" and "smooth-pasting" conditions, \cite{Kwok2022, Chiarella2009} which say that the American option value and its first derivative in $z$ are continuous at the exercise boundary.

It is known that for $\beta < 0$ there is a positive probability for $y_t$ to hit zero, \cite{LinetskyMendozza2010}. That raises a natural question: could SOFR go negative? In principle, yes. Indeed, very low SOFRs around 0.01\% have already been observed at the market. As explained in \cite{DerivativeLogic2020}, "the Federal’s aggressive moves to maintain liquidity in the repo markets are driving it ever downward. As a borrower, if you had financed a floating rate loan that used SOFR as the floating rate index, you would be loving it right now. If you are the lender, however, you would be seeking an adjustment to your lending rate that better reflects the true cost of credit (trying to widen the credit spread).  Nowadays, this is seen as a lender's “hack” of SOFR in the event of a blowup in credit markets. Should lenders have to hack a secured rate (SOFR) that was believed to be a lender’s best replacement for a flawed unsecured rate (LIBOR) in the first place? That is a question for another day".

Having this in mind, in this paper we assume that SOFR is always non-negative, and, hence, set
\begin{equation}
y_l(t) = [ -\barR^*(t)]^+.
\end{equation}
Here we use the function $x^+ = \max(x,0)$ because $y_t$ is assumed to be non-negative under the CEV model). Accordingly, we need an additional boundary condition for $P(t, -\barR^*(t), z)$ for which the natural choice would be
\begin{equation} \label{bc1}
P(t, y_l(t), z) = K.
\end{equation}
Now, comparing this with \eqref{bc0}, to make them consistent for the corner case $ z = z_B(t, y_l(t))$ we must assume that
\begin{equation} \label{Corner}
 z_B(t, y_l(t)) = 0.
 \end{equation}

The main idea of pricing American options by considering only the continuation region where the exercise boundary and the option price can be found simultaneously by solving a certain linear integral Volterra equation of the second type (LIVESK) together with a nonlinear algebraic equation, has been proposed and implemented in a series of papers \cite{CarrItkin2020jd, ItkinMuravey2024jd, Itkin2024jd}. Here, the main trick to proceeding with a semi-analytical solution of the problem in \cref{PDE,tc0,bc0,bc1} lies in the utilization of this idea as applied to Asian-American options.

It is known that for equity Asian options, when the underlying stock price follows the GBM with constant coefficients, the corresponding pricing PDE (in a sense of \eqref{PDE}) can be transformed into a one-dimensional PDE with respect to a new variable $x$, \cite{RogersShi1995, Alziary1997}
\begin{equation}
y_t \mapsto x_t = \frac{1}{y_{t_0}} \left[ K - \frac{1}{t-t_0}\int_{t_0}^t y(\tau) \mu(d\tau)\right],
\end{equation}
\noindent where $\mu$ is a probability measure with density $\rho(t)$ on the interval $[t_0, T]$, and $\rho(t) = 1/T$ for average rate options. However, in the case considered in this paper, this or a similar transformation does not give rise to a one-dimensional PDE.

\subsection{Finding the exercise boundary and the option price}

To solve the problem in \cref{PDE,tc0,bc0,bc1} we combine two approaches already elaborated in our papers. The first one consists of solving \eqref{PDE} without the killing drift term in the variable $z$, since for the remaining PDE this was done in \cite{CarrItkinMuravey2020, ItkinMuravey2024jd}. Then, the solution of the full PDE in \eqref{PDE} can be obtained by using a generalized Duhamel's principle, as explained in detail in \cite{ItkinMuravey2024jd, Itkin2024jd}.

\subsubsection{Solution of the problem in \cref{PDE,tc0,bc0,bc1} without last two terms in \eqref{PDE}} \label{homogBessel}

A similar problem for the Equity American Call option has already been solved in \cite{ItkinMuravey2024jd}. Notice, that the PDE for the Call option can be transformed to \eqref{PDE} by substituting $K \mapsto -K, S \mapsto -S$. Using the approach of \cite{ItkinMuravey2024jd}, the solution of \cref{PDE,tc0,bc0,bc1} can be obtained as follows.

As shown in \cite{CarrItkinMuravey2020}, by a set of transformations \eqref{PDE} without the last two terms can be transformed to the Bessel PDE
\begin{proposition} \label{prop1}
The PDE in \eqref{PDE} (without the last two terms) can be transformed to
\begin{equation} \label{Bess}
\fp{u}{\tau} =\frac{1}{2} \sop{u}{x} + \frac{b}{x} \fp{u}{x},
\end{equation}
\noindent where $b$ is some constant, $u = u(\tau, x, z)$ is the new dependent variable, and $(\tau, x, z)$ are the new independent variables. The \eqref{Bess} is the PDE associated with the one-dimensional Bessel process, \cite{RevuzYor1999}
\begin{equation} \label{BesProc}
d X_t = d W_t  + \frac{b}{X_t} dt.
\end{equation}

\begin{proof}[{\bf Proof}]
The full proof can be found in \cite{CarrItkinMuravey2020}. In short, the transformation is done in two steps. First, we use the change of variables
\begin{equation} \label{tr1}
y = \left(-\chi \beta \right)^{-1/\beta}, \qquad P(t,y) \to \hat{u}(\phi,\chi,z), \qquad \phi = \int_t^T \sigma^2(k) dk,
\end{equation}
\noindent which reduces \eqref{PDE} to the form
\begin{align} \label{reduction1}
\fp{\hat{u}}{\phi} &= \frac{1}{2} \sop{\hat{u}}{\chi} + \left[ f(t) \chi + \frac{b}{\chi} \right] \fp{\hat{u}}{\chi}, \\
f(t) &= - \beta \frac{\alpha  (t)}{\sigma^2(t)}, \qquad t = t (\phi), \qquad b =\frac{1 + \beta}{2\beta}, \nonumber
\end{align}
\noindent and then the other change of variables
\begin{align} \label{tr2}
x &= \chi F(\phi), \quad \hat{u}(t,\chi,z) \to u(\tau, x, z), \quad \tau  = \int_0^{\phi} F^2(k) dk, \quad F(\phi) = \exp \left[ \int_0^\phi f(k) dk\right],
\end{align}
The PDE in \eqref{Bess} should be solved in the domain
\begin{equation*}
\Omega: (\tau, x. z) \in [0, \tau(t_0)] \times [x_l(\tau), \infty) \times [z_B(\tau, x), \infty), \qquad
x_l(t) = - \frac{F(\phi(t))}{\beta} y^{-\beta}_l(t), \quad t = t(\tau),
\end{equation*}
\noindent subject to the terminal condition
\begin{equation} \label{tc0cev}
u(0, x, z) = 0,
\end{equation}
\noindent and the boundary conditions
\begin{align} \label{bc0cev}
u(\tau,x \uparrow \infty, z) &= 0, \qquad u(\tau,x_l(t(\tau)), z) = K, \\
u(\tau,x, z_B(\tau, x)) &\equiv f^+(\tau,x) = K - z_B(\tau,x). \nonumber
\end{align}

Also, since $z_B(0,x) = K$ (again, see discussion in \cite{ItkinMuravey2024jd}), we have $f^+(0,x) = 0$.
\end{proof}
\end{proposition}

As discussed in \cite{CarrItkinMuravey2020}, the above formulas are valid at  $-1 < \beta < 0$. However, if $0 < \beta < 1$, the right boundary moves to zero while the left boundary moves to some negative value $- 1/(\beta y_l(\tau)^\beta)$.
Therefore, in this case it is convenient to redefine $y \to \tilde{y} = -y$. This also redefines $x \to \tilde{x} = -x$. Then, the domain of definition for $\tilde{x}$ becomes $\tilde{x} \in [0, \tilde{x}_r(\tau)],\ \tilde{x}_r(\tau) = - x_l(\tau)$. Also, in case $0 < \beta < 1$ the PDE in \eqref{Bess} keeps the same form in the $\tilde{x}$, \cite{CarrItkinMuravey2020}.

Similar to \cite{ItkinMuravey2024jd}, we can reduce the problem in \eqref{Bess}, \eqref{tc0cev}, \eqref{bc0cev} to that with the homogeneous terminal and boundary conditions by making a change of variables\footnote{It is assumed that $x_l(\tau) \ne 0$, but the case $x_l(\tau) = 0$ can be handled in the same way while all the below formulae become much simpler.}
\begin{equation} \label{trHomo}
U(\tau,x,z) = u(\tau, x,z) - \frac{x_l(\tau) }{x} K
\end{equation}
This yields
\begin{align} \label{wEq}
\fp{U}{\tau} &= \frac{1}{2} \sop{U}{x} + \frac{b}{x} \fp{U}{x} + \lambda(\tau,x), \qquad \lambda(\tau,x) = \frac{K}{x} \left[ \frac{(1 - b) }{x^2} x_l(\tau) - x'_l(\tau) \right], \\
U(0,x,z) &= - \frac{x_l(0)}{x} K, \qquad U(\tau,\infty,z) = U(\tau, x_l(\tau),z) = 0. \nonumber
\end{align}
Then, Duhamel's principle can be applied to solve this problem if the solution to the same problem but with no source term $\lambda(\tau,x)$ is known\footnote{Generalization of Duhamel's principle as applied to problems with moving boundaries is described in \cite{ItkinMuravey2024jd, Itkin2024jd}. }.

\paragraph{The Weber-Orr Theta function.} In \cite{CarrItkinMuravey2020, ItkinMuravey2024jd} when solving a similar one-dimensional problem for the Call option in the domain $\Omega_y: y \in [0, y_B(t)]$ (so, no dependence on variable $z$), a {\it periodic} solution of the Bessel equation at $\Omega_y$ was constructed. The authors named it the {\it Bessel Theta function} $\Theta_{\nu}(\theta, x_1, x_2)$ - a new special function, which is an analog of the Jacobi Theta function, \cite{mumford1983tata} for the heat equation, and defined as follows
\begin{equation} \label{thetaBess}
	\Theta_{\nu}(\theta, x_1, x_2)   = 2 (x_1 x_2)^{|\nu|} \sum_{n = 1}^\infty e^{- \mu_n^2 \theta^2}  \frac{\Jnum( \mu_n x_1) \Jnum(\mu_n x_2)}{J_{|\nu| + 1}^2 (\mu_n)}.
\end{equation}
Here $J_\nu(x)$ is the Bessel function of the first kind, $\nu = 1/(2\beta) < 0$, since $\beta < 0$, $\mu_n$ is an ordered sequence of positive zeros of the Bessel function $\JnuM(\mu)$:
\begin{equation*}
	J_{|\nu|}(\mu_n) = J_{|\nu|}(\mu_m) = 0, \quad \mu_n > \mu_m > 0, \quad n >m.
\end{equation*}
As shown in \cite{ItkinMuravey2024jd}, the sum in the definition of $\Theta_\nu(...)$ usually quickly converges due to the exponential term, so taking even about 10 terms in the sum provides an accuracy of the order of $10^{-16}$.

A similar approach advocated in this paper for the domain $\Omega^\infty_y: y \in [y_B(t), \infty)$ gives rise to another new special function, which we call as the {\it Weber-Orr Theta function} $\varTheta_{|\nu|}(\tau, v, w, a)$ and define it as follows
\begin{equation}  \label{WeberOrrTheta}
\varTheta_{|\nu|}(\tau, v, w, a) = \int_0^\infty  e^{-\frac{p^2}{2}\tau} \frac{W_{|\nu|}(p, v, a) W_{|\nu|}(p, w, a)}{V_{|\nu|}(p, a)} p\, dp.
\end{equation}
Here, $W_{|\nu|}(p, v, a)$ - the kernel of the Weber-Orr transform, \cite{bateman1953higher}, and the function $V_{|\nu|}(p, v)$ are defined as follows
\begin{align} \label{WeberOrrKernel_def}
W_{|\nu|}(p, v, a) &= \JnuM(p v) \Ynu(p a) -  \Ynu(p v) \JnuM(p a), \\
V_{|\nu|}(p, a) &= \JnuM^2(p\, a) + \Ynu^2(p\, a), \nonumber
\end{align}
\noindent where $\Ynu(z)$ denotes the Bessel function of the second kind (also known as the Neumann or Weber function) which is linearly independent of $\Jnu(z)$. This construction, despite not being recognized as a new special function, was introduced in \cite{CarrItkinMuravey2020}. As mentioned there, the definitions in \eqref{WeberOrrKernel_def} are generalizations of the Pythagorean and Angle Sum identities for trigonometric functions to the case of cylinder functions $\JnuM$ and $\Ynu$.  The functions $W_{|\nu|}(p,v,a)$ as functions of the second argument $v$ also form an orthogonal basis in the space ${\mathbb C}(\Omega^\infty_y)$ for all $\tau > 0$.

The Weber-Orr Theta function can also be thought of as an inverse Weber-Orr transform $W^{-1}_{\nu, \mu}[\hat{g};p,v]$, see \cite{Titchmarsh:75,Gorshkov2019} among others. This representation reads
\begin{equation}
\varTheta_{|\nu|}(\tau, v, w, a) = W^{-1}_{|\nu|, |\nu|}[\hat{g};p,v], \qquad g(p,\tau, w, a) = e^{-\frac{p^2}{2}\tau}W_{|\nu|}(p, w, a).
\end{equation}

\begin{figure}
\centering
\fbox{\includegraphics[width=\textwidth]{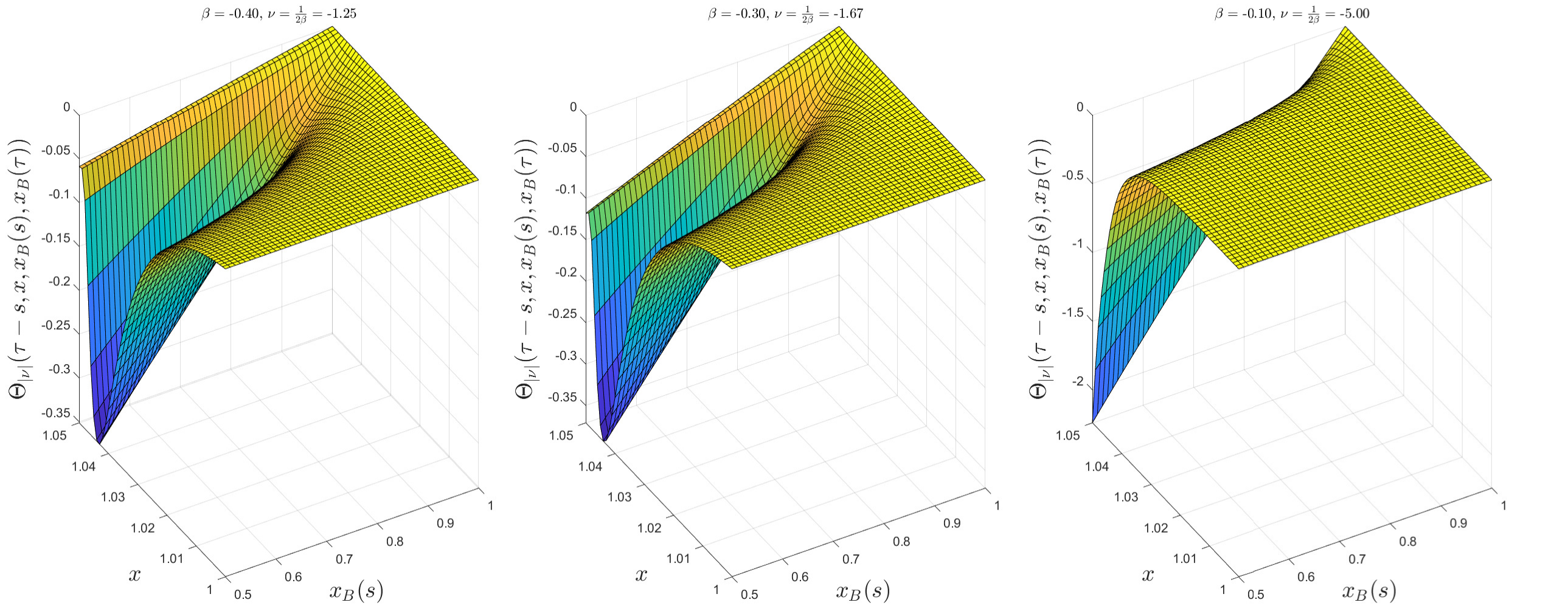}}
\caption{Function $\varTheta_{|\nu|}(\tau - s, x, x_l(s), x_l(\tau))$ for three values of $\beta = -0.4, -0.3, -0.1$ with
$\tau = 0.5,\, s \in (0, \tau],\, x_l(\tau) = 1.0,\, x_l(s) = x_l(\tau) - s,\, x(s) = x_l(\tau) + 0.1 s$.}
\label{FigTheta}
\end{figure}

As an illustration in Fig.~\ref{FigTheta} we present a 3D plot of $\varTheta_{|\nu|}(\tau - s, x, x_l(s), x_l(\tau))$ (see \eqref{uBessSolHomo}) for three values of $\beta = -0.4, -0.3, -0.1$.

In what follows, we will also need the first derivative of $\varTheta_{|\nu|}(\tau, v, w, a)$ on the second argument $v$ which we further denote as $\varTheta'_{|\nu|}(\tau, v, w, a)$. From \eqref{WeberOrrTheta} and \eqref{WeberOrrKernel_def} we obtain
\begin{equation}  \label{WeberOrrThetaPrime}
\varTheta'_{|\nu|}(\tau, v, w, a) = \int_0^\infty  p\, e^{-\frac{p^2}{2}\tau} \frac{W'_{|\nu|}(p, v, a) W_{|\nu|}(p, w, a)}{V_{|\nu|}(p, a)} dp,
\end{equation}
\noindent where
\begin{alignat}{2} \label{WeberOrrKernel_defPrime}
W'_{|\nu|}(p, v, a) &= p\JnuM'(p v) \Ynu(p a) -  p\Ynu'(p v) \JnuM(p a),
\qquad& \JnuM'(x) &= J_{|\nu|-1}(x) - \frac{|\nu|}{x}\JnuM(x).
\end{alignat}

In what follows,  we need several important properties of the Weber-Orr Theta functions. First, it can be directly checked that
\begin{equation} \label{woThetaProp1}
\varTheta_{|\nu|}(\tau, v, w, v) = 0.
\end{equation}
Regarding other properties, see \cref{betam1}.

\paragraph{Solution of the problem.} Luckily, the homogeneous problem in \eqref{wEq} and, respectively, in \cref{Bess,tc0cev,bc0cev} has been solved in \cite{CarrItkinMuravey2020} (all the details and the corresponding discussion can be found in that paper). The solution reads (with $z$ being a dummy variable)
\begin{align}  \label{uBessSolHomo}
u(\tau, x, z) &=  \frac{x_l(\tau) }{x} K + x^{-\nu} \Bigg\{ \frac{1}{2}\int_{0}^\tau x_l^{\nu + 1}(s) \Psi(s, x_l(s)) \varTheta_{|\nu|}(\tau - s, x, x_l(s), x_l(\tau) ) ds \\
&- K x_l(0) \int_{x_l(0)}^\infty \xi^{\nu} \varTheta_{|\nu|}(\tau, x, \xi, x_l(\tau) ) d\xi \Bigg\}. \nonumber
\end{align}
Note, that the Weber-Orr Theta function $\varTheta_{|\nu|}(\tau, v, w, a)$ is also a {\it periodic} solution of the Bessel equation at the semi-infinite domain $\Omega_y^\infty$.

Another, not yet defined term in \eqref{uBessSolHomo}, $\Psi(\tau, x_l(\tau))$ is the Put option Delta (the gradient of the solution) at the moving boundary $x_l(\tau)$
\begin{align} \label{lPsi}
\Psi(\tau, x_l(\tau),z) = \left. \fp{u(\tau,x,z)}{x} \right|_{x = x_l(\tau)}.
\end{align}
It can be found explicitly by differentiating both parts of \eqref{uBessSolHomo} on $x$, substituting $x \to x_l(\tau)$ and taking into account that $\varTheta_{|\nu|}(\tau - s, x_l(\tau), x_l(s), x_l(\tau) ) = 0$. This yields
\begin{align} \label{grCEV}
\Psi(\tau, x_l(\tau)) &=  - \frac{K}{x_l(\tau)} + x_l(\tau)^{-\nu} \Bigg\{ \frac{1}{2} \int_{0}^\tau x_l^{\nu + 1}(s) \Psi(s, x_l(s)) \varTheta'_{|\nu|}(\tau - s, x_l(\tau), x_l(s), x_l(\tau) ) ds \\
&- K x_l(0) \int_{x_l(0)}^\infty \xi^{\nu} \varTheta'_{|\nu|}(\tau, x_l(\tau), \xi, x_l(\tau) ) d\xi \Bigg\}. \nonumber
\end{align}
Thus, the gradient $\Psi(\tau, x_l(\tau))$ solves a LIVESK. This equation can be efficiently solved numerically, in more detail, see \cite{ItkinLiptonMuraveyBook}. To emphasize, as follows from \eqref{grCEV}, $\Psi(\tau, x_l(\tau))$ doesn't depend on $z$.

\paragraph{Adding a source term.}
Similar to \cite{ItkinMuravey2024jd}, the additional source term in \eqref{wEq} can be taken into account by using a generalized Duhamel's principle, \cite{ItkinMuravey2024jd, Itkin2024jd}. This yields
\begin{align}  \label{uBessSolInHomo}
u(\tau, x, z) &=  \frac{x_l(\tau) }{x} K + x^{-\nu} \Bigg\{ \frac{1}{2}\int_{0}^\tau x_l^{\nu + 1}(s) \Psi(s, x_l(s)) \varTheta_{|\nu|}(\tau - s, x, x_l(s), x_l(\tau) ) ds \\
&- K x_l(0) \int_{x_l(0)}^\infty \xi^{\nu} \varTheta_{|\nu|}(\tau, x, \xi, x_l(\tau) ) d\xi \Bigg\}
+ \int_0^\tau \int_{x_l(s)}^\infty \lambda(s, \xi) \varTheta_{|\nu|}(\tau - s, x, \xi, x_l(\tau) ) d\xi ds. \nonumber
\end{align}

Accordingly, instead on \eqref{uBessSolHomo} one now has the following equation for $\Psi(\tau, x_l(\tau),z)$
\begin{align} \label{grCEV1}
\Psi(\tau, x_l(\tau)) &=  - \frac{K}{x_l(\tau)} + x_l(\tau)^{-\nu} \Bigg\{ \frac{1}{2} \int_{0}^\tau x_l^{\nu + 1}(s) \Psi(s, x_l(s)) \varTheta'_{|\nu|}(\tau - s, x_l(\tau), x_l(s), x_l(\tau) ) ds \\
&- K x_l(0) \int_{x_l(0)}^\infty \xi^{\nu} \varTheta'_{|\nu|}(\tau, x_l(\tau), \xi, x_l(\tau) ) d\xi \Bigg\}
+ \int_0^\tau \int_{x_l(s)}^\infty \lambda(s, \xi) \varTheta'_{|\nu|}(\tau - s, x_l(\tau), \xi, x_l(\tau) ) d\xi ds. \nonumber
\end{align}

\subsubsection{Solution of the full problem in \cref{PDE,tc0,bc0,bc1}} \label{secSolAsian}

Again, by using Duhamel's principle for PDEs with moving boundaries, \cite{ItkinMuravey2024jd, Itkin2024jd}, we can show that the last two terms $[\barR^*(t) + y] \fp{P}{z} - [\barR^*(t) + y] P$ in \eqref{PDE} can be added as an additional term to the source term $\lambda(\tau,x,z)$ in \eqref{wEq}. Then, with this change taken into account, \eqref{uBessSolInHomo} still holds, but with a redefined $\lambda(\tau,x,z)$.

In more detail, by making the same chain of transformations as described in \cref{homogBessel}, but now accounting for the extra terms in \eqref{PDE}, instead of \eqref{Bess} we obtain\footnote{The expressions for $B(\tau,x), C(\tau,x,z)$ are obtained by again using Proposition~\ref{prop1}, but now with all terms of the PDE in \eqref{PDE} included.}
\begin{align} \label{BessSource}
\fp{u}{\tau} &= \frac{1}{2} \sop{u}{x} + \frac{b}{x} \fp{u}{x} + C(\tau,x,z) \fp{u}{z} - B(\tau,x) u, \\
B(\tau,x) &=  \frac{1}{\sigma^2(t)}e^{- 2 \int_0^{\phi(t) } f(k) \, dk} \left[ \barR^*(t) + y(t, x) \right], \quad y(t, x) = x^{-1/\beta} \left[-\beta/F(\phi(t)) \right]^{-1/\beta}, \quad t = t(\tau), \nonumber \\
C(\tau,x,z) &=  \frac{1}{(t - t_0) \sigma^2(t)}e^{- 2 \int_0^{\phi(t) } f(k) \, dk} \left[ \barR^*(t) + y(t, x) - z\right],  \nonumber
\end{align}
\noindent with the same terminal and boundary conditions as in \eqref{tc0cev}, \eqref{bc0cev}.

To switch to the problem with homogeneous boundary and terminal conditions, as in \eqref{wEq}, we make the same change to the dependent variable as in \eqref{trHomo}, and obtain a new expression for the source term in \eqref{wEq} which is now denoted as $\Lambda(\tau,x,z)$
\begin{align} \label{newLambda}
\Lambda(\tau, x,z) &= \lamIn(\tau, x) + C(\tau,x,z) \fp{u}{z} - B(\tau,x) u, \\
\lamIn(\tau, x) &= \lambda(\tau, x) - \frac{x_l(\tau) }{x} K B(\tau,x). \nonumber
\end{align}

Accordingly, we intend to use the solution in \eqref{uBessSolInHomo} and apply the generalized Duhamel's principle. When doing so, one needs to take into account that the Green function of the problem depends on variables $(t,x,z)$ while the PDE in \eqref{BessSource} doesn't contain the second derivative in $z$ (hence, this PDE is degenerate in this variable). Since from the definition of $z$ it follows that $z \in [0,\infty]$, its two-dimensional Green's function could be represented in the form $G(t,x,z) = G^{(x)}(t,x) G^{(z)}(t,z)$. Applying the method of images \cite{Howison1995} to the semi-infinite interval yields
\begin{equation} \label{images}
G^{(z)}(t,z | T, \zeta) = \delta(z-\zeta) - \delta(z+\zeta).
\end{equation}
Thus, using this result, from \eqref{uBessSolInHomo} and the generalized Duhamel's principle, we obtain
\begin{align}  \label{uBessFull}
u(&\tau, x, z) =  \frac{x_l(\tau) }{x} K \\
&+ x^{-\nu} \Bigg\{ \frac{1}{2}\int_{0}^\tau x_l^{\nu + 1}(s) \Psi(s, x_l(s),z) \varTheta_{|\nu|}(\tau - s, x, x_l(s), x_l(\tau) ) ds
- K x_l(0) \int_{x_l(0)}^\infty \xi^{\nu} \varTheta_{|\nu|}(\tau, x, \xi, x_l(\tau) ) d\xi \Bigg\} \nonumber \\
&+ \int_0^\tau \int_{x_l(s)}^\infty \Big\{ \lamIn(s, \xi) + C(s,\xi,z) u_z(s,\xi, z) - B(s,\xi) u(s,\xi, z) \Big\} \varTheta_{|\nu|}(\tau - s, x, \xi, x_l(\tau) ) d\xi ds. \nonumber
\end{align}
Here, formally speaking, the gradient $\Psi(s, x_l(s),z)$ has to be a function of $z$ as well. The \eqref{uBessFull} is not a closed form solution to the problem in \cref{PDE,tc0,bc0,bc1}, but rather an integral Volterra equation of the second kind with respect to $u(\tau,x,z)$ if $z_B(\tau,x)$ and $\Psi(s, x_l(s),z)$ are known. However, e.g., $z_B(\tau,x)$ for American options is obviously not known in advance and has to be computed simultaneously with the option price. Nevertheless, by using a standard approach of the GIT method, \cite{ItkinLiptonMuraveyBook}, one may derive a couple of auxiliary equations to make this solution fully determined.

\paragraph{Solution for $\bm{z_B(\tau,x)}$.}

First, we substitute $z = z_B(\tau,x)$ into both parts of \eqref{uBessFull}. By the smooth fit principle, \cite{Kwok2022, Chiarella2009} the option value and the first derivative in $z$ are continuous at the exercise boundary (while the second derivative experiences a jump). Hence, the value of $u_z(\tau, x, z_B(\tau))$ has to be understood here as $\left. u_z(\tau, x, z)\right|_{z \to z_B(\tau,x)^+}$, i.e., as the limit of $P_z(\tau, x, z)$ in the continuation region at $z \to z_B(\tau, x)$. Accordingly, we obtain
\begin{align}  \label{uBessFull2}
K\Big( 1 &- \frac{x_l(\tau) }{x} \Big) - z_B(\tau,x) = x^{-\nu} \Bigg\{ \frac{1}{2}\int_{0}^\tau x_l^{\nu + 1}(s) \Psi(s, x_l(s), z_B(s,x)) \varTheta_{|\nu|}(\tau - s, x, x_l(s), x_l(\tau) ) ds \nonumber \\
&- K x_l(0) \int_{x_l(0)}^\infty \xi^{\nu} \varTheta_{|\nu|}(\tau, x, \xi, x_l(\tau) ) d\xi \Bigg\}
+ \int_0^\tau \int_{x_l(s)}^\infty \varTheta_{|\nu|}(\tau - s, x, \xi, x_l(\tau) ) \\
&\cdot \Big\{ \lamIn(s, \xi) - B(s,\xi) [K - z_B(s,\xi)]  - C(s,\xi,z_B(s,\xi)) \Big\} d\xi ds. \nonumber
\end{align}

In addition to $z_B(\tau,x)$, this equation  also depends on a yet unknown gradient $\Psi(\tau, x_l(\tau), z_B(\tau,x))$.
Therefore, to proceed, we  differentiate both parts of \eqref{uBessFull} by $x$ and set $x = x_l(\tau)$ to obtain
\begin{align}  \label{uBessFullPsi}
\Psi(&\tau, x_l(\tau), z) =  - \frac{K}{x_l(\tau)} + x^{-\nu}_l(\tau) \Bigg\{ - K x_l(0) \int_{x_l(0)}^\infty \xi^{\nu} \varTheta'_{|\nu|}(\tau,x_l(\tau), \xi, x_l(\tau) ) d\xi \\
&+ \frac{1}{2}\int_{0}^\tau x_l^{\nu + 1}(s) \Psi(s, x_l(s),z) \varTheta'_{|\nu|}(\tau - s, x_l(\tau), x_l(s), x_l(\tau) ) ds  \Bigg\} \nonumber \\
&+ \int_0^\tau \int_{x_l(s)}^\infty \Big\{ \lamIn(s, \xi) + C(s,\xi,z) u_z(s,\xi, z) - B(s,\xi) u(s,\xi, z) \Big\} \varTheta'_{|\nu|}(\tau - s, x_l(\tau), \xi, x_l(\tau) ) d\xi ds. \nonumber
\end{align}
For the particular case $z = z_B(\tau,x)$, \eqref{uBessFullPsi} takes the form\footnote{When differentiating, we take into account \eqref{woThetaProp1}, so no derivative of $\Psi(s, x_l(s),z_B(s,x))$ on $x$ appears in the RHS of \eqref{uBessFullPsiB}.}
\begin{align}  \label{uBessFullPsiB}
\Psi(&\tau, x_l(\tau), z_B(\tau, x)) =  - \frac{K}{x_l(\tau)} + x^{-\nu}_l(\tau) \Bigg\{ - K x_l(0) \int_{x_l(0)}^\infty \xi^{\nu} \varTheta'_{|\nu|}(\tau,x_l(\tau), \xi, x_l(\tau) ) d\xi \\
&+ \frac{1}{2}\int_{0}^\tau x_l^{\nu + 1}(s) \Psi(s, x_l(s), z_B(s,x)) \varTheta'_{|\nu|}(\tau - s, x_l(\tau), x_l(s), x_l(\tau) ) ds \Bigg\} \nonumber \\
&+ \int_0^\tau \int_{x_l(s)}^\infty \Big\{ \lamIn(s, \xi) - B(s,\xi) [K - z_B(s,\xi)]  - C(s,\xi,z_B(s,\xi))  \Big\} \varTheta'_{|\nu|}(\tau - s, x_l(\tau), \xi, x_l(\tau) ) d\xi ds. \nonumber
\end{align}

 The \cref{uBessFullPsiB,uBessFull2} can be solved together. Suppose we introduce a discrete grid $\mathbb{T}$ in the time $\tau$ as $\mathbb{T}: 0, \tau_1, ... \tau_N = \tau(t_0)$ which contains $N+1$ nodes (in the simplest case, the grid could be uniform). At every node $\tau_i, \ 1 \le i \le N$ the values of $\Psi(\tau_i, x_l(\tau_i), z_B(\tau,x)), z_B(\tau,x)$ can be found as follows:
 \begin{enumerate}[label=(\alph*)]
    \item Start with some initial guess for $z_B(\tau,x) = z^{(0)}_B(\tau,x)$;
    \item Substitute this guess  into \eqref{uBessFullPsiB} to obtain a LIVESK for $\Psi(\tau_i, x_l(\tau_i), z^{(0)}_B(\tau,x))$. Solve it.
    \item Substitute this solution into \eqref{uBessFull2} to obtain a nonlinear algebraic equation for $z_B(\tau,x)$. Solve it to obtain the next approximation, $z^{(1)}_B(\tau,x)$.
    \item  Go to item (b) and repeat steps (b)-(d) until convergence is reached.
 \end{enumerate}

 \paragraph{Solution for  $\bm{u(\tau, x, z)}$.}

 Since $z_B(\tau,x)$ and $\Psi(s, x_l(s),z_B(\tau,x) $ are already known from the previous step, one can replace some terms in the integrals in curly braces in \eqref{uBessFull} and \eqref{uBessFullPsi}  with their representation from \eqref{uBessFull2}. This yields
\begin{align}  \label{uBessFullNew}
\bar{u}(\tau, x, z) &=   \int_0^\tau \int_{x_l(s)}^\infty \left[  C(s,\xi,z) \bar{u}_z(s,\xi, z) + C(s,\xi,z_B(s,\xi)) - B(s,\xi)  \bar{u}(s,\xi, z) \right] \varTheta_{|\nu|}(\tau - s, x, \xi, x_l(\tau) ) d\xi ds \nonumber \\
&+  \frac{1}{2} x^{-\nu} \int_{0}^\tau x_l^{\nu + 1}(s) [\Psi(s, x_l(s),z) - \Psi(s, x_l(s),z_B(s,x) ] \varTheta_{|\nu|}(\tau - s, x_l(\tau), x_l(s), x_l(\tau) ) ds,  \\
\bar{u}(\tau,x, z) &= u(\tau,x,z) -  (K  - z_B(\tau,x)), \qquad \bar{u}(\tau,x, z_B(\tau,x)) = 0. \nonumber
\end{align}
This equation can be solved together with \eqref{uBessFullPsi} re-written in terms of $\bar{u}(\tau, x, z)$
\begin{align}  \label{uBessFullPsiNew}
\Psi(&\tau, x_l(\tau), z) =  - \frac{K}{x_l(\tau)} + x^{-\nu}_l(\tau) \Bigg\{ - K x_l(0) \int_{x_l(0)}^\infty \xi^{\nu} \varTheta'_{|\nu|}(\tau,x_l(\tau), \xi, x_l(\tau) ) d\xi \\
&+ \frac{1}{2}\int_{0}^\tau x_l^{\nu + 1}(s) \Psi(s, x_l(s),z) \varTheta'_{|\nu|}(\tau - s, x_l(\tau), x_l(s), x_l(\tau) ) ds  \Bigg\} \nonumber \\
&+ \int_0^\tau \int_{x_l(s)}^\infty \Big\{ \blamIn(s, \xi) + C(s,\xi,z) \bar{u}_z(s,\xi, z) - B(s,\xi) \bar{u}(s,\xi, z) \Big\} \varTheta'_{|\nu|}(\tau - s, x_l(\tau), \xi, x_l(\tau) ) d\xi ds, \nonumber \\
\blamIn(\tau, x) &= \lamIn(\tau, x) - B(\tau,x) [K -  z_B(\tau, x)]. \nonumber
\end{align}

To emphasize, \cref{uBessFullNew,uBessFullPsiNew} are not standard Volterra equations of the second kind (because they also contain the term $\bar{u}_z(\tau, x, z)$ under the integral in the RHS. However, since the value of $\bar{u}(\tau, x, z)$ at the exercise boundary $z_B(\tau,x)$ is already known and equal to zero, in the numerical scheme of integration we can approximate this derivative with the finite difference\footnote{This will be relaxed in \cref{finalization}, and no FD scheme would be necessary to solve this equation.}
\begin{equation} \label{discret}
\bar{u}_z(\tau_i, x, z_k) ) = \frac{1}{\Delta z} \left[ \bar{u}(\tau_i, x, z_k) - \bar{u}(\tau_i, x, z_k - \Delta z) \right] + O(\Delta z).
\end{equation}
In other words, we solve this problem by additionally introducing some discrete (e.g., uniform) grid $\mathbb{Z}: z_B(\tau, x), z_B(\tau, x) + \Delta z,\dots,z-\Delta z, z$ in the $z$ space, with $\Delta z$ being the grid step. With this approximation,\footnote{Higher order approximations in $\Delta z$ can be constructed in a similar way.} \cref{uBessFullNew,uBessFullPsiNew} becomes a system of two {\it LIVESK}, where, however, the integration is done only in $(\tau, x)$ space, so $z$ stays as a dummy variable. Once $\bar{u}(\tau_i, x, z_B(\tau_i,x) + \Delta z)$ is found, one can proceed to getting $\bar{u}(\tau_i, x, z_B(\tau_i,x) + 2\Delta z)$ in the same way, and so on until the point $z$..

By definition in \eqref{z}, the option price we need is just $u(\tau_i, x, z(t_0))$. However, due to the presence of $\bar{u}_z(\tau, x, z)$ in the RHS of \eqref{uBessFullNew}, it can be obtained only sequentially on a grid $\mathbb{Z}: z \in [z_B(\tau,x), \barR^*(t_0) + y(t_0,x) - z(t_0)]$ (assuming $z(t_0) > z_B(\tau,x)$).

The \cref{uBessFullNew,uBessFullPsiNew} can be numerically solved at the 3D grid $\mathbb{T} \bigotimes \mathbb{X} \bigotimes \mathbb{Z}$,  as follows.
\begin{enumerate}[label=(\alph*)]
    \item Start with the initial guess, $\Psi(s, x_l(s),z) = \Psi(s, x_l(s),z_B(\tau,x)$, since $\Psi(s, x_l(s),z_B(\tau,x)$ is already known.
    \item Substitute this guess  into \eqref{uBessFullNew} to obtain a LIVESK for $\bar{u}(\tau_i, x,z)$. Solve it.
    \item Substitute this solution into \eqref{uBessFullPsiNew} to obtain another LIVESK for $\Psi(s, x_l(s),z)$. Solve it to obtain the next approximation.
    \item  Go to item (b) and repeat steps (b)-(d) until convergence is reached.
\end{enumerate}

It is important that during these iterations only the source terms of these LIVESK change while the matrices of the LHS don't. Therefore, iterations don't influence the order of computational complexity $O(2 N^2 M L)$ where $L$ is the number of nodes in the $\mathbb{Z}$ grid.

Alternatively, \cref{uBessFullNew,uBessFullPsiNew} can be solved as a 2D LIVESK, see, e.g., \cite{Saeed2011} among others.

\subsubsection{Pricing Asian options in the whole region} \label{asianE}

To price Asian options with the European style payoff using the method proposed in above, one needs to take into account that the payoff function in \eqref{tc0} now doesn't vanish outside of the continuation region. In other words, instead of \eqref{tc0} we need to use the terminal condition
\begin{equation} \label{tc0E}
	P(T, y, z) = (K - z(T))^+.
\end{equation}
Accordingly, the terminal condition in \eqref{tc0cev} also changes to
\begin{equation} \label{tc0cevE}
u(0, x, z) = (K - z(T))^+.
\end{equation}
Then the problem in \eqref{wEq} also acquires a modified terminal condition
\begin{equation}
U(0,x,z) = (K - z(T))^+ - \frac{x_l(0)}{x} K.
\end{equation}
Then, with allowance for \eqref{images} and this new terminal condition, in the second term in the curly braces in \eqref{uBessFull} the multiplier $K x_l(0)$ should be replaced with
\begin{equation} \label{trE}
K x_l(0) \to K x_l(0) - x (K - z)^+.
\end{equation}
The same change should be propagated to \eqref{uBessFullPsiNew} while it doesn't affect the equations for $z_B(\tau,x)$ in \cref{uBessFull2,uBessFullPsiB}.

Finally, since for the European style Asian option there is no such a notion as the exercise boundary, integration in $z$ could be done starting from $z = \infty$ where the Put option value should vanish.

Once the price of the Asian option with the European payoff has been determined, the corresponding price of the Asian option with the American payoff style can be found as described in \cref{1mAmerican}.

\subsection{Intermediate results} \label{IntRes}

To the best of our knowledge, so far there has been no development on semi-analytical pricing of Asian options with the American style exercise, even for time-homogeneous models. In this section, by utilizing the GIT approach (proposed in \cite{CarrItkin2020jd} and described in detail in \cite{ItkinLiptonMuraveyBook}) in general, and its particular peculiarities for semi-analytical pricing of the American options, \cite{CarrItkin2020jd, ItkinMuravey2024jd, Itkin2024jd} we have shown how this can be done for these options written on the SOFR (as the underlying), where the SOFR itself follows a time-inhomogeneous CEV model.

The solution is obtained in a few steps:
\begin{enumerate}
	\item First, the system \cref{uBessFullPsiB,uBessFull2} is solved to obtain the exercise boundary $z_B(\tau, x)$. This can be done by using a discrete grid $\mathbb{T}$ in $\tau \in [0, \tau(t_0)]$ space, and also another discrete grid $\mathbb{X}$ in $x$.. Since $x \in [x_l(\tau), \infty)$, we truncate the upper boundary to some large value $x_r(\tau) < \infty$, so $x \in [x_l(\tau), x_r(\tau)]$. Then the grid is constructed as $x_l(\tau), x_1(\tau) + \Delta x, ... x_l(\tau) + M \Delta x = x_r(\tau)$ which contains $M+1$ nodes (again, in the simplest case, the grid could be uniform). The integrals in these equations are discretized by using some quadratures.

Given $z_B(\tau,x_j), \, x_j,\, 1 < j \le M$, \eqref{uBessFullPsiB} (which is a LIVESK for $\Psi(\tau, x_l(\tau),  z_B(\tau,x)_j)$ )  becomes a linear system of algebraic equations (for every $0 \le s \le \tau$) with a triangular matrix, see \cite{ItkinLiptonMuraveyBook}. The numerical complexity of this solution is $O(N^2)$. Moreover, this system is the same for every $x_j,\, 1 < j \le M$ except only the source term (the last term in the RHS of \eqref{uBessFullPsiB}). Therefore, the total complexity of the solution remains $O(N^2)$. Solving the nonlinear algebraic equations \eqref{uBessFull2} is of the complexity $O(N^2 M)$. Assuming that a typical number of iterations required for the nonlinear solver to converge is $I$, the total complexity of this step is $O(N^2 M I)$.

\item At this step, the Put option price $\bar{u}(\tau_i, x, z)$ is found, by first discretizing $\bar{u}_z(\tau_i, x, z)$ according to \eqref{discret}, and then solving the system \cref{uBessFullNew,uBessFullPsiNew}, which after discretization is a system of two LIVESK with respect to $\bar{u}(\tau_i, x, z)$ and $\Psi(\tau_i, x_l(\tau), z)$. To solve it, we need a 3D grid $\mathbb{T} \bigotimes \mathbb{X} \bigotimes \mathbb{Z}$. The numerical complexity of solving this system is $O(N^2 M L)$.

\end{enumerate}

The main difference between this result and those obtained by using the GIT method for pricing American options, \cite{ItkinMuravey2024jd, Itkin2024jd} lies in the fact, that \eqref{uBessFull2} is not an {\it explicit} representation of the option price as in \cite{ItkinMuravey2024jd, Itkin2024jd}, but rather a linear integral equation with respect to the option price. This is because of the extra term $[\barR^*(t) + y] \fp{P}{z}$ in \eqref{PDE}. From this perspective, here the term "semi-analytical" might sound a bit confusing. However, as shown \cref{simplification}, those LIVESKs can be further simplified by using the properties of the Weber-Orr Theta function at $\tau \to 0$ proved in Propositions~\ref{prop2}-\ref{prop3}. This gives to a true semi-analytical (and recurrent) solution for both the option value and the gradient when solving these LIVESKs sequentially in time. To underline, on one hand, this can be treated as an efficient numerical method of solving these equations. On the other hand, by splitting the whole time interval $s$ into the two: $s \in [0, \tau(t)] = [0, \tau^-(t)] \bigcup (\tau^-(t), \tau(t)]$ and using those properties of the Weber-Orr Theta function at the second interval, this semi-analyticity can become transparent in general.

Overall, when solving LIVESK, the numerical complexity of our method is $O(N^2 M H)$, where $H = \max(I,L)$. This can be compared with that of, e.g., a FD approach. It is worth mentioning that accurate pricing of Asian options requires a special technique due to the presence of only the first derivative in $z$ in \eqref{PDE}. In more detail, see \cite{ZvanForsythVetzal1997}. Approximating all the derivatives in \eqref{PDE} with the second order in time and space steps, the total complexity of the solution for the European Asian option is about $O(N M_1 L_1)$. Despite $M_1 \approx M$, however, $L_1 \gg H$ since the FD scheme needs all points in both continuation and exercise regions, while in our case we deal only with the continuation region. For the American-Asian option, since the exercise boundary is not explicitly known (this is in contrast to our method), one needs to additionally apply some kind of penalty method, \cite{Halluin2004, ZvanForsythVetzal1997}. This requires several iterations to maintain the required accuracy. Also, the accuracy of our method can be easily increased by using higher-order quadratures, e.g., the Simpson integration rule with the fourth order of approximation in the grid step, \cite{ItkinLiptonMuraveyBook}. This reduces the complexity of our method to $O(N^2 M^{1/2} H)$ providing the same accuracy as before, while still having $L_1 \gg H$. At the same time, for the FD scheme reaching a higher order of approximation is not easy and also increases the complexity of the numerical solution, \cite{ItkinBook}.

It is even possible to combine these two methods, meaning you can first find the exercise boundary $z_B(\tau, x)$ with the complexity $O(N^2 M I)$, and then in the continuation region solve \eqref{PDE} with the European payoff by using the FD method with the complexity $O(N M_1 L_1)$. Hence, the total complexity becomes $O(N \cdot \max(N M I,  M_1 L_1))$.

Another important advantage of the GIT is the computation of option Greeks, \cite{ItkinLiptonMuraveyBook}. The main point is that the Greeks, i.e., derivatives of the solution, can also be expressed semi-analytically by differentiating the solution (or, both parts of the corresponding LIVESK) on some parameter or independent variable of the model. This results in a new integral equation for the corresponding option Greek, which can be solved in a similar way.. In contrast, e.g., for the FD method, the Greeks can be found only by bumping the corresponding parameters. This might create various numerical issues, especially in the regions where the solution is not smooth enough (for instance, close to the exercise boundary), while for this version of the GIT method all Greeks can be calculated by re-running only the last step in the above procedure.

This is especially important for American options since, in contrast to the option value and Delta, the option Gamma and Vega experience a jump at the exercise boundary, meaning their values vanish in the exercise region,  but not in the continuation region, \cite{ItkinMuravey2024jd}. An accurate computation of this jump is always a problem for the FD methods, and they need to use a version of the penalty method to accurately resolve it, \cite{Halluin2004}.

At the end, let us mention that the solution also depends on $y_{t_0}$ such that $\barr(t_0) + R^*(t_0) + y_{t_0}$ is the instantaneous SOFR forward price at  time $t_0$. It can be found as this is described in \cref{American}. In other words, $y_{t_0} = f(t_0, \bar{y}, Q)$, where $\bar{y} = y_t|_{t = \tins}$, and $Q$ is the maturity of the corresponding ZCB, see \cref{solForward}.

\subsection{Simplifications and finalization} \label{simplification}

\subsubsection{Particular values of the CEV $\beta$} \label{betam1}

Since the CEV model covers various known models by making a particular choice of the parameter $\beta$, in this section we consider $\beta = -1$. The corresponding model is known as the Hull-White model, \cite{Hull:1990a,andersen2010interest}, which in our case has all parameters of the model to be time-dependent.

As under the Hull-White model the underlying has a Gaussian distribution, various formulae derived in the previous sections become either more transparent or significantly simplified in this case. First of all, this is true for the Weber-Orr Theta function which in the case $\beta = -1$ transforms into a combination of exponents. Accordingly, the following results hold:
\begin{proposition} \label{prop2}
For the Hull-White model ($\beta=-1$) the Weber-Orr Theta function has the following properties
\begin{align} \label{WOThetaProp}
\varTheta_{|\nu|}(\tau, v, w, a) &= \frac{1}{\sqrt{v w}} \frac{1}{\sqrt{2 \pi \tau }} \left[e^{-\frac{(v-w)^2}{2 \tau }} - e^{-\frac{(v+w - 2a)^2}{2 \tau }}\right], \\
\lim_{\tau \to 0} \varTheta_{|\nu|}(\tau, v, w, a) &= \frac{1}{\sqrt{v w}} \left[ \delta(v-w) - \delta(v + w - 2a) \right], \nonumber \\
\varTheta'_{|\nu|}(\tau, v, w, v) &= \frac{2(w-v)}{\sqrt{v w}} \frac{1}{\sqrt{ 2 \pi \tau^3}} e^{-\frac{(v-w)^2}{2 \tau }}, \quad \Re(\tau) > 0, \nonumber \\
\lim_{\tau \to 0} \varTheta'_{|\nu|}(\tau, v, w, v)  &= \frac{2}{\sqrt{v w}}\delta'(v-w),
\qquad \lim_{\tau \to 0} \lim_{v \to w} \varTheta'_{|\nu|}(\tau, v, w, v)  = 0. \nonumber
\end{align}
\noindent where $\delta(x)$ is the Dirac Delta function.

\begin{proof}[{\bf Proof}]
Since for the Hull-White model $\beta = -1$, from the definition of $\nu = 1/(2 \beta)$ given after Fig.~\ref{FigTheta} we have $\nu = -1/2$. Accordingly, the Bessel functions with index $1/2$ are expressed via trigonometric functions, or imaginary exponents, \cite{as64}. Substituting  this representation into the definition of the Weber-Orr Theta function in \eqref{thetaBess} and integrating, we obtain the first line in \eqref{WOThetaProp}.

Differentiation of this line by $v$ and substitution $ a \to v$ gives rise to the third line in \eqref{WOThetaProp}. It can also be re-written in the form
\begin{equation*}
\varTheta'_{|\nu|}(\tau, v, w, v) = \frac{2}{\sqrt{v w}} \frac{\partial }{\partial v} \left[ \frac{1}{\sqrt{2 \pi  \tau }} e^{-\frac{(v-w)^2}{2 \tau }} \right].
\end{equation*}
The expression in  square brackets is the heat kernel, \cite{Kartashov2001}. Then, the known result is that at $\tau \to 0$ the heat kernel tends to the Dirac Delta function. Based on this fact, the second and the fourth lines of \eqref{WOThetaProp} follow.
\end{proof}
\end{proposition}

\subsubsection{Extension for an arbitrary $\beta$}

The main question to immediately and naturally ask based on results obtained in \cref{betam1} would be whether Proposition~\ref{prop2} is also valid in the general case of an arbitrary $\beta$, i.e., whether the properties listed in  this proposition are common properties of the Weber-Orr Theta function. Proposition~\ref{prop3} provides some guidance to answer it positively.

\begin{proposition} \label{prop3}
For general $\nu = 1/(2\beta)$ with $\beta \in [-1,1]$ the second and fourth lines of proposition~\ref{prop2} still hold.

\begin{proof}[{\bf Proof}]
By definition of the Weber-Orr Theta function in \eqref{WeberOrrTheta}, at $\tau \to 0$ we have
\begin{equation}  \label{WeberOrrTau0}
\lim_{\tau \to 0} \varTheta_{|\nu|}(\tau, v, w, a) = \int_0^\infty  \frac{W_{|\nu|}(p, v, a) W_{|\nu|}(p, w, a)}{V_{|\nu|}(p, a)} p\, dp.
\end{equation}

According to Weber's integral theorem, \cite{Titchmarsh1921}, if $v > a > 0$ and $f(t)$ is of bounded variation in a neighborhood of $t = v$, then the following identity holds
\begin{align} \label{Wident}
\int_0^\infty  \frac{W_{|\nu|}(p, v, a) }{V_{|\nu|}(p, a)} p\, dp \int_a^\infty W_{|\nu|}(p, t, a) t f(t) dt =
\frac{1}{2}[ f(x+0 + f(x-0)].
\end{align}
On the other hand, it is always true, that $\varTheta_{|\nu|}(\tau, v, w, v) = 0$.

Therefore, let us set
\begin{equation*}
f(t) = \frac{1}{\sqrt{t }} [\delta(t-w) - \delta(2a - t - w)],
\end{equation*}
\noindent  where $w > a > 0$, and substituting it into \eqref{Wident}. This yields
\begin{align} \label{Wident2}
\int_0^\infty  p \frac{W_{|\nu|}(p, v, a) W_{|\nu|}(p, w, a)}{V_{|\nu|}(p, a)}  \sqrt{w} dp =
\frac{1}{\sqrt{v}}[\delta(v - w) - \delta(v + w - 2a)].
\end{align}
Here, in the RHS of \eqref{Wident} we used the oddness of the Dirac Delta function, and in the LHS - the fact that
\begin{equation*}
\int_a^\infty W_{|\nu|}(p, t, a) \sqrt{t}\,  \delta(2a - (t - w) dt = 0,
\end{equation*}
\noindent since $ w > a$. Combining \cref{WeberOrrTau0,Wident2} we obtain the second line of \eqref{WOThetaProp} for an arbitrary $\beta$. Also, we have
\begin{equation*}
\lim_{a \to v} \lim_{\tau \to 0} \varTheta_{|\nu|}(\tau, v, w, a)  =
\lim_{\tau \to 0} \lim_{a \to v} \varTheta_{|\nu|}(\tau, v, w, a) = 0,
\end{equation*}
\noindent as it should be. Note, that the function $\frac{1}{\sqrt{v}}\delta(v - w)$ is of bounded variation if $v \ne w$.

Now, the last identity in the fourth line of \eqref{WOThetaProp} follows by first, differentiating the second line by $v$ and then taking the limit $a \to v$.

\end{proof}
\end{proposition}

\subsubsection{Simplifications of  the final results} \label{finalization}

It turns out, that these simplifications not only look attractive, but also drastically affect the way the solution in the previous sub-sections could be obtained. This is related to \eqref{uBessFullPsiB} and \eqref{uBessFull2}.

Indeed, in \cref{secSolAsian} instead of a semi-analytical expression for $\Psi(\tau, x_l(\tau), z) $ we derived two LIVESK, \eqref{uBessFullPsiB} and \eqref{uBessFull2}, that should be solved numerically. But, in the first integral on $s$ in the RHS of \eqref{uBessFullPsiB}, based on Proposition~\ref{prop3}, at $ s \to \tau$ we obtain $\varTheta'_{|\nu|}(\tau - s, x_l(\tau), x_l(s), x_l(\tau) ) \to 0$. Accordingly, in the second integral we have
\begin{align}
\lim_{s \to \tau} & \int_{x_l(s)}^\infty \Big\{ \lamIn(s, \xi) - B(s,\xi) [K -  z_B(s,\xi)] - C(s,\xi,z_B(s,\xi)) ]   \Big\} \varTheta'_{|\nu|}(\tau - s, x_l(\tau), \xi, x_l(\tau) ) d\xi \\
&= \int_{x_l(\tau)}^\infty \Big\{ \lamIn(\tau, \xi) - B(\tau,\xi) [K - z_B(\tau,\xi)]  - C(\tau,\xi,z_B(\tau, \xi)) ]   \Big\} \frac{2}{\sqrt{x_l(\tau) \xi}}\delta'(\xi - x_l(\tau)) d\xi = 0, \nonumber
\end{align}
Thus, all terms in the RHS of \eqref{uBessFullPsiB} corresponding to $s = \tau$ vanish. But, since \cref{uBessFullPsiB,uBessFull2} are solved sequentially in time at the grid $\mathbb{T}$, all the remaining terms under the integrals in the RHS of these equations, corresponding to $s < \tau$, are already known by the time $\tau$. And, due to the vanished kernels at $s \to \tau$, the term $\Psi(\tau, x_l(\tau), z)$ disappears from the RHS of \cref{uBessFullPsiB,uBessFullPsiB}. Thus, \cref{uBessFullPsiB,uBessFullPsiB} are no more LIVESK, but rather a recurrent in time semi-analytical solution for $z_B(\tau,x))$ and $\Psi(\tau, x_l(\tau), z_B(\tau,x))$. Same is true for \eqref{uBessFullNew,uBessFullPsiNew}.

In more detail, suppose the solution of the problem has already been found up to the point $\tau_i \in \mathbb{T}: 0, \tau_1, ... \tau_N = \tau(t_0)$, such that $1 < i < N$. For the shortness of notation let us assume up to the end of this section that $\tau = \tau_{i+1}, \tau_m = \tau_i$. With allowance for the properties of the Weber-Orr Theta function listed in Proposition~\ref{prop3}, from  \eqref{uBessFull2} we have
\begin{align}  \label{uBessFul3}
&\qquad \qquad K\Big( 1 - \frac{x_l(\tau) }{x} \Big) -  z_B(\tau,x)  - \frac{1}{2} \Delta \tau \calI(\tau) = A(\tau, x) + \frac{1}{2} \Delta \tau \calI_1(\tau_m),  \\
A(\tau,x) &=  x^{-\nu} \Bigg\{ \frac{1}{2}\int_{0}^{\tau_m} x_l^{\nu + 1}(s) \Psi(s, x_l(s), z_B(s,x)) \varTheta_{|\nu|}(\tau - s, x, x_l(s), x_l(\tau) ) ds \nonumber \\
&- K x_l(0) \int_{x_l(0)}^\infty \xi^{\nu} \varTheta_{|\nu|}(\tau, x, \xi, x_l(\tau) ) d\xi \Bigg\}
+ \int_0^{\tau_m} \int_{x_l(s)}^\infty \varTheta_{|\nu|}(\tau - s, x, \xi, x_l(\tau) ) \calL_B(s,\xi) d\xi ds, \nonumber \\
\calL_B(s,\xi) &= \lamIn(s, \xi)  - B(s,\xi) [K - z_B(s,\xi)] - C(s,\xi,z_B(s,\xi)),  \nonumber \\
\calI_1(\tau_m) &= \frac{1}{2} x^{-\nu} x_l^{\nu + 1}(\tau_m) \Psi(\tau_m, x_l(\tau_m), z_B(\tau_m,x)) \varTheta_{|\nu|}(\Delta \tau, x, x_l(\tau_m), x_l(\tau) ) \nonumber \\
&+ \int_{x_l(\tau_m)}^\infty \varTheta_{|\nu|}(\Delta \tau, x, \xi, x_l(\tau) ) \calL_B(\tau_m,\xi) d\xi, \qquad \Delta \tau = \tau - \tau_m, \nonumber \\
\calI(\tau) &= \int_{x_l(\tau)}^\infty  \frac{1}{\sqrt{x \xi}} \left[ \delta(x-\xi) - \delta(x + \xi - 2 x_l(\tau)) \right] \calL_B(\tau, \xi) d\xi \nonumber \\
&= \frac{1}{x} \calL_B(\tau,x) - \frac{1}{\sqrt{x x_l(\tau)}} \calL_B(\tau,2 x_l(\tau)-x) \Ind_{x = x_l(\tau)} = \frac{\Ind_{x \ne x_l(\tau)}}{x} \calL_B(\tau,x), \nonumber
\end{align}
\noindent because, based on \eqref{Corner}, $z_B(\tau, x_l(\tau)) = 0$. In \eqref{uBessFul3} we used a trapezoidal quadrature for the approximation of the integral $\int_{\tau_m}^\tau \ldots ds$, however, this can be easily relaxed.

Since $\calL_B(\tau,x)$ is linear in $z_B(\tau,x)$, \eqref{uBessFul3} can be trivially solved to get a closed form representation
\begin{align} \label{zbFinal}
z_B(\tau, x) &= \frac{K \left( 1 - \frac{x_l(\tau) }{x} \right) - A(\tau,x) - \frac{1}{2} \Delta \tau \Bigg[ \calI_1(\tau_m)
+ \frac{\Ind_{x \ne x_l(\tau)}}{x} \Big( \lamIn(\tau, x) - B(\tau,x) K - C(\tau,x,0) \Big) \Bigg]}{1 + \frac{\Ind_{x \ne x_l(\tau)}}{2 x} \Delta \tau [B(\tau,x) + Q(\tau)]}, \nonumber \\
Q(\tau) &= \frac{1}{(t - t_0) \sigma^2(t)}e^{- 2 \int_0^{\phi(t) } f(k) \, dk}, \qquad t = t(\tau).
\end{align}

Similarly, from \eqref{uBessFullPsiB} we have
\begin{align}  \label{uBessFullPsiBrec}
\Psi(&\tau, x_l(\tau), z_B(\tau, x)) =  - \frac{K}{x_l(\tau)} + x^{-\nu}_l(\tau) \Bigg\{ - K x_l(0) \int_{x_l(0)}^\infty \xi^{\nu} \varTheta'_{|\nu|}(\tau,x_l(\tau), \xi, x_l(\tau) ) d\xi \\
&+ \frac{1}{2}\int_{0}^{\tau_m} x_l^{\nu + 1}(s) \Psi(s, x_l(s), z_B(s,x)) \varTheta'_{|\nu|}(\tau - s, x_l(\tau), x_l(s), x_l(\tau) ) ds \Bigg\} \nonumber \\
&+ \int_0^{\tau_m} \int_{x_l(s)}^\infty \calL_B(s,\xi) \varTheta'_{|\nu|}(\tau - s, x_l(\tau), \xi, x_l(\tau) ) d\xi ds  \nonumber \\
&+ \frac{1}{2} \Delta \tau \Bigg[ \frac{1}{2} x^{-\nu}_l(\tau) x_l^{\nu + 1}(\tau_m) \Psi(\tau_m, x_l(\tau_m), z_B(\tau_m,x)) \varTheta'_{|\nu|}(\Delta \tau, x_l(\tau), x_l(\tau_m), x_l(\tau) ) \nonumber \\
&+ \int_{x_l(\tau_m)}^\infty \calL_B(\tau_m,\xi) \varTheta'_{|\nu|}(\Delta \tau, x_l(\tau), \xi, x_l(\tau) ) d\xi   \Bigg]. \nonumber
\end{align}
Thus, instead of solving a system of two LIVESK in \eqref{uBessFullNew} at every time step $\tau_i$, actually we have obtained a closed form  (recurrent) solution for $z_B(\tau,x)$ and $\Psi(\tau, x_l(\tau), z_B(\tau ,x))$.

\paragraph{The exercise boundary at $\mathbf{t=t_0}$.} It is well-known that the PDE in \eqref{PDE} has a singularity at $t=t_0$. Usually, the standard approach is to further assume that $z(t_0) = \barR^*(t_0) + y$ to resolve it. This trick works well for numerical methods that don't utilize an explicit value of the exercise boundary, but rather implicitly compare the European and intrinsic values of the option at every computational point. However, when trying to find a closed form solution for $z_B(\tau(t_0),x)$ as we do it in this paper, this immediately brings a problem.

Indeed, the standard approach simply removes the first derivative $P_z$ from the PDE at $t=t_0$, so the PDE becomes  independent of $z$, i.e., becomes one-dimensional in the space state. If we do that in \eqref{zbFinal}, it can be seen that $z_B(\tau),x)$ will experience a jump at $t \to t_0$ and tend to $z_B(\tau(t_0),x) = K$. Thus, the option price for the OTM (out of the money) options will vanish, which is unrealistic and contradicts known numerical results. Another reason for that is that the standard assumption sets $z(t_0) = \barR^*(t_0) + y$, but not $z_B(\tau(t_0),x) = \barR^*(t_0) + y$.
On the other hand, since known numerical methods utilize an implicit check for the exercise condition, such an approach could be questioned at $t \to t_0$, subject to a bigger and, to a certain extent, uncontrolled error.

Another way to resolve this problem could be as follows. One can look at the limit of \eqref{zbFinal} at $t \to t_0$.  It it is easy to check from \eqref{zbFinal} that in this case  $z_B(\tau(t_0),x) \to \barR^*(t_0) + y$. Again, this condition means that the exercise boundary experiences a jump when $t \to t_0$. But since in the continuation region $K < \barR^*(t_0) + y$, the intrinsic value of the option also becomes zero.

Thus, both of the described approaches don't resolve this problem. Therefore, in this paper we redefine the original problem by setting that $z_B(\tau(t_0),x)$ can be determined by interpolating $z_B(\tau,x)$ over time given the already computed values of $z_B(\tau_j,x), \ \tau_j \in [\tau_1 = 0,\ldots,\tau_{N-1} = \tau(t_0) - \Delta \tau]$. This assumption, despite not being justified theoretically, eliminates jumps at $t \to t_0$ and provides plausible values of the Asian-American Put option prices, see \cref{experiments}.

\paragraph{Simplification of the solution for the option price.} Closely looking at the RHS of \eqref{uBessFullNew}, one can recognize that, by using Proposition~\ref{prop3},  the first term in the limit $s \to \tau$ can also be computed explicitly to yield
\begin{align} \label{uHW}
\lim_{s \to \tau} & \int_0^\tau \int_{x_l(s)}^\infty D(s,\xi,z) \varTheta_{|\nu|}(\tau - s, x, \xi, x_l(\tau) ) d\xi ds
= \frac{1}{\sqrt{x x_l(\tau)}}  \times    \\
& \int_{x_l(\tau)}^\infty D(\tau,\xi,z) \left[ \delta(x-\xi) - \delta(x + \xi - 2x_l (\tau)) \right] d\xi =  \frac{\Ind_{x \ne x_l(\tau)}}{x} D(\tau, x, z), \nonumber \\
D(s,\xi, z) &= - B(s,\xi) \bar{u}(s,\xi, z) + C(s,\xi,z) \bar{u}_z(s,\xi, z) + C(s,\xi,z_B(s,\xi)). \nonumber
\end{align}

This means that \eqref{uBessFullNew} also is not anymore a LIVESK, but rather a recurrent solution for $u(\tau,x,z)$.  Indeed, combining \eqref{uBessFullNew} and \eqref{uHW} yields
\begin{align}  \label{uBessFullNew2}
\bar{u}(\tau, x, z) &=  \calA(\tau,x,z) + \frac{\Ind_{x \ne x_l(\tau)}}{2 x}\Delta \tau D(\tau, x, z), \\
\calA(\tau,x,z) &= \int_0^{\tau_m} \int_{x_l(s)}^\infty D(s,\xi, z)  \varTheta_{|\nu|}(\tau - s, x, \xi, x_l(\tau) ) d\xi ds \nonumber \\
&+  \frac{1}{2} x^{-\nu} \int_{0}^{\tau_m} x_l^{\nu + 1}(s) [\Psi(s, x_l(s),z) - \Psi(s, x_l(s),z_B(s,x) ] \varTheta_{|\nu|}(\tau - s, x_l(\tau), x_l(s), x_l(\tau) ) ds,  \nonumber \\
&+ \frac{1}{2} \Delta \tau \Bigg\{ \int_{x_l(\tau_m)}^\infty D(\tau_m,\xi, z) \varTheta_{|\nu|}(\Delta \tau, x, \xi, x_l(\tau) ) d\xi  \nonumber \\
&+  \frac{1}{2} x^{-\nu} x_l^{\nu + 1}(\tau_m) [\Psi(\tau_m, x_l(\tau_m),z) - \Psi(\tau_m, x_l(\tau_m),z_B(\tau_m,x)) ] \varTheta_{|\nu|}(\Delta \tau, x_l(\tau), x_l(\tau_m), x_l(\tau) ) \Bigg\},  \nonumber
\end{align}
\noindent where for $\Psi(s, x_l(s),z)$ we have from \eqref{uBessFullPsiNew}
\begin{align}  \label{uBessFullPsiRec}
\Psi(&\tau, x_l(\tau), z) =  - \frac{K}{x_l(\tau)} + x^{-\nu}_l(\tau) \Bigg\{ - K x_l(0) \int_{x_l(0)}^\infty \xi^{\nu} \varTheta'_{|\nu|}(\tau,x_l(\tau), \xi, x_l(\tau) ) d\xi \\
&+ \frac{1}{2}\int_{0}^{\tau_m} x_l^{\nu + 1}(s) \Psi(s, x_l(s), z) \varTheta'_{|\nu|}(\tau - s, x_l(\tau), x_l(s), x_l(\tau) ) ds \Bigg\} \nonumber \\
&+ \int_0^{\tau_m} \int_{x_l(s)}^\infty \calL(s,\xi)  \varTheta'_{|\nu|}(\tau - s, x_l(\tau), \xi, x_l(\tau) ) d\xi ds  + \frac{1}{2} \Delta \tau \Bigg[ \frac{1}{2} x^{-\nu}_l(\tau) x_l^{\nu + 1}(\tau_m) \Psi(\tau_m, x_l(\tau_m), z) \nonumber \\
&\times  \varTheta'_{|\nu|}(\Delta \tau, x_l(\tau), x_l(\tau_m), x_l(\tau) )
+ \int_{x_l(\tau_m)}^\infty \calL(\tau_m, \xi) \varTheta'_{|\nu|}(\Delta \tau, x_l(\tau), \xi, x_l(\tau) ) d\xi   \Bigg], \nonumber \\
\calL(\tau,x) &= \lamIn(\tau,x) - B(\tau,x) \bar{u}(\tau,x, z) + C(\tau,x,z)\bar{u}_z(\tau,x, z). \nonumber
\end{align}

The \eqref{uBessFullNew2} could be treated as a linear ordinary differential equation (ODE) with the dependent variable $\bar{u}(\tau,x,z)$ and the independent variable $z$ if re-written in the form
\begin{align} \label{ODE}
\left[1 + a(x) B(\tau,x) \right] \bar{u}(\tau,x, z_j) &- a(x) \bar{u}_z(\tau,x, z_j) \left[ C(\tau,x,0) - Q(\tau) z_j \right] = \calA(\tau,x,z_j) + a(x) C(\tau,x,z_B(\tau,x)), \nonumber \\
a(x) &=  \frac{\Ind_{x \ne x_l(\tau)}}{2 x}\Delta \tau, \qquad z_j \in [z_B(\tau,x), z], \quad j=1,\ldots,N_z.
\end{align}
It can be solved subject to the boundary condition $\bar{u}(\tau,x,z_B(\tau,x)) = 0$, which follows from \eqref{bc0cev}, and the definition of $\bar{u}(\tau,x,z)$ in \eqref{uBessFullNew}. This yields
\begin{align} \label{ODEsol}
\bar{u}(\tau,x, z_j) &= - \int_{z_B(\tau,x)}^{z_j}
\left(\frac{C(\tau,x,0) - Q(\tau) k }{C(\tau,x,0) - Q(\tau) z_j }\right)^p
\frac{\calA(\tau,x,k) + a(x) C(\tau,x,z_B(\tau,x))}{C(\tau,x,0) - Q(\tau) k} dk, \qquad x \ne x_l(\tau) \nonumber \\
\bar{u}(\tau,x_l(\tau), z_j) &= \calA(\tau,x_l(\tau),z_j), \qquad p = \frac{1 + a(x) B(\tau,x)}{a(x) Q(\tau)}.
\end{align}
Note, that in case $\calA(\tau,x,z)$ doesn't depend on $z$, the integral in \eqref{ODEsol} can be taken explicitly to yield
\begin{align}
\bar{u}(\tau,x, z_j) &=  \frac{\calA(\tau,x) + a(x) C(\tau,x,z_B(\tau,x))}{ Q(\tau)} \left [ 1 - \frac{1}{p} \left(
\frac{C(\tau,x, z_B(\tau,x))}{C(\tau,x,z_j)} \right)^p \right].
\end{align}
By the definition of $a(x)$ in \eqref{ODE}, it is small and $a(x) \to 0$ when $x \to \infty$. Accordingly, the coefficient $p$
demonstrates the same behavior. Since
\begin{equation*}
\lim_{p \to 0} \frac{1}{p} (a/b)^p = - \log(a/b),
\end{equation*}
\noindent the solution in \eqref{ODEsol} is well-behaved if $C(\tau,x, z_B(\tau,x)) > C(\tau,x,z)$ at large $x$, but otherwise could become negative. Therefore, we additionally need to set
\begin{equation}
\bar{u}(\tau,x, z_j) = 0, \quad \frac{1}{p} \left( \frac{C(\tau,x, z_B(\tau,x))}{C(\tau,x,z_j)} \right)^p > 1, \quad
\textrm{or} \quad \frac{C(\tau,x, z_B(\tau,x))}{C(\tau,x,z_j)} > \log\left( p^{1/p} \right).
\end{equation}
Obviously, $\log\left( p^{1/p}\right) < 0$ since $p < 1$. But $C(\tau,x,z_j)$ can also be negative. Also, as this directly follows from \eqref{ODE}, at small $a(x)$ the solution of this ODE tends to $\bar{u}(\tau,x, z_j) = \calA(\tau,x_l(\tau),z_j)$.

The results in \eqref{ODEsol} and below show that we don't need any FD approximation of the first derivative $\bar{u}_z(\tau,x, z)$ to obtain the final solution (despite the fact that the  FD approximation of the first derivative in $z$ can alternatively be used to solve \eqref{ODE}). However, we still need to numerically compute the integrals in variables $\tau, x,z$.

Accordingly, from \eqref{ODE} we have
\begin{align} \label{ODEsolDer}
\bar{u}_z(\tau,x, z) &=  \frac{ \left[ 1 + a(x) B(\tau,x) \right] \bar{u}(\tau,x, z)  - \calA(\tau,x,z)  -  a(x) C(\tau,x,z_B(\tau,x))}{a(x) \left[ C(\tau,x,0) - Q(\tau) z \right] }.
\end{align}
However, at $a(x) \to 0$ the value of $\bar{u}_z(\tau,x, z)$ becomes undetermined. But, since $u(\tau,x,z) \to 0$ when $x \to \infty$, it is expected that in this limit $\bar{u}_z(\tau,x, z) \to -1$.

The recurrent solution in variables $z, x$ can be obtained on a grid $\mathbb{X} \bigotimes \mathbb{Z}$ using the fact, that in the RHS of \eqref{uBessFullNew2} all terms are already known at the time $\tau$ because i) the option prices $u(\tau_m,x,z) $ have been already computed at the previous time step $\tau_m < \tau$, and ii) the same is true for the gradients $\Psi(s, x_l(s), z)$ since $s \le \tau_m$. Thus, we have a true semi-analytical solution to the problem.

To emphasize,  this solution is constructed to deliver the European Put option price only in the continuation region, so the terminal condition in \eqref{tc0} has been used. That was done to achieve our ultimate goal of deriving a LIVESK for the exercise boundary. However, as shown in \cref{asianE}, by a minor modification \eqref{trE}, it can also provide the European Put option price for the entire domain.

\section{Pricing American options on futures} \label{American}

In this section we discuss how to price options on futures in the interval $t \in [S_{\mathrm{ins}}, SR_{\mathrm{start}}]$, where, in fact, they are American options written on a forward rate.

This is a much simpler problem than that discussed in \cref{Asian}. Accordingly, it can be solved by using the results already obtained in \cref{Asian} and \cite{ItkinMuravey2024jd}. Indeed, in \cite{ItkinMuravey2024jd} American options written on ZCB $F(t,y,Q)$ with the maturity $Q$, where the underlying interest rate $y$ follows the time-dependent CEV model, have already been  considered. A semi-analytical representation for the Call option price at the interval $y \in [0, y_B(t)]$, where $y_B(t)$ is the exercise boundary in the $y$ space \footnote{This boundary is uniquely translated into $F_B(t,y_B(t),Q)$ - the exercise boundary in the $F$ space.}, has also been obtained by using the GIT method. As, by definition, \cite{BM2006}, the instantaneous forward rate $f(t,Q)$ reads
\begin{equation} \label{forwardDef}
	f(t,y, Q) = - \fp{}{Q} \log F(t,y,Q),
\end{equation}
\noindent one can pick this connection and reuse the result in \cite{ItkinMuravey2024jd}. However, since here we consider an American Put option with a different continuation region $y \in [y_B(t), \infty)$, the result in \cite{ItkinMuravey2024jd} should be combined with the result of \cref{Asian}. Below, we describe this approach in more detail.

Again, we assume that the dynamics of $y_t$ is described by the CEV model as in \cref{Asian}. Using the same argument as in \cite{ItkinMuravey2024jd}, assume that the exercise boundary $y_B(t)$ is known. Then, in the continuation region, the pricing problem for the American Put $P(t,y)$ is equivalent to pricing Down-and-Out barrier option at the domain $\Omega: y \in [y_B(t), \infty) \times t \in [0,T_f]$, where $T_f$ is the maturity of this American option written on futures (in the next section we will discuss how to set $T_f$). Hence, the American Put price $P(t,y)$ in the continuation region solves the PDE\footnote{In this section $y$ refers to $y_t|_{t = \tins}$. }
\begin{equation} \label{PDEP}
\fp{P}{t} - \alpha(t) y \fp{P}{y} + \frac{1}{2} \sigma^2(t) y^{2 \beta+2}\sop{P}{y} - [\barR^*(t) + y] P = 0, \qquad(t, y) \in \mathbb{R}_+ \times [y_B(t), \infty),
\end{equation}
\noindent subject to the terminal condition
\begin{equation} \label{tchw}
	P(T_f,y) = \left(K - f(T_f, y, Q)\right)^+ = 0,
\end{equation}
\noindent the boundary condition at the moving boundary
\begin{equation} \label{bc1hw}
	P(t,y_B(t)) = K - f(t, y_B(t), Q),
\end{equation}
\noindent and the other boundary condition at $y \to \infty$. Since, based on \eqref{bc1-42}, in this limit the forward price tends to infinity, this yields
\begin{equation} \label{bc2hw}
	P(t,y)\Big|_{y \to \infty}  = 0.
\end{equation}

Further, we use Proposition~\ref{prop1} to transform this problem into the following one:
\begin{align} \label{BessAmer}
\fp{u}{\tau} &= \frac{1}{2} \sop{u}{x} + \frac{b}{x} \fp{u}{x} - B(\tau,x) u, \\
u(0, x) = 0, \qquad u(\tau, x \uparrow \infty) &= 0, \qquad u(\tau, x_B(\tau)) \equiv f^+(\tau) = K - f(\tau,x_B(\tau),Q), \nonumber
\end{align}
\noindent and then make another change of variables (similar to \eqref{trHomo})
\begin{equation} \label{trHomoAmer}
	U(\tau,x) = u(\tau, x) - \frac{x_B(\tau)}{x} f^+(\tau),
\end{equation}
\noindent that. together with Duhamel's principle, yields
\begin{align} \label{wEqF}
\fp{U}{\tau} &= \frac{1}{2} \sop{U}{x} + \frac{b}{x} \fp{U}{x} + \lambda(\tau,x) - g(\tau,x) \left[ U + \frac{x_B(\tau)}{x} f^+(\tau)\right], \\
\lambda(\tau,x) &= \frac{x_B(\tau) f^+(\tau)}{x}\left[ \frac{1 - b}{x^2} - \fp{}{\tau}\log[x_B(\tau) f^+(\tau)] \right], \qquad g(\tau,x) = B(\tau,x), \nonumber \\
U(0,x) &= - \frac{x_B(0)}{x} f^+(0), \qquad U(\tau, x \uparrow \infty) = U(\tau, x_B(\tau)) = 0. \nonumber
\end{align}

The solution of this problem has already been obtained in \cref{secSolAsian} and can be utilized here if an explicit dependence of $f(\tau, x_B(\tau),Q))$ on $x_B(\tau, Q)$ is known. Therefore, at this point, we need to return to the problem in \eqref{PDE-42}, \eqref{termZCB-42}, \eqref{bc1-42} to determine such a dependence.

Note, that the terminal condition for the continuation region in \eqref{tchw} is sufficient to determine the exercise boundary. Then, to find the European option price in the whole region (including the exercise region), a modified boundary condition
\begin{equation}
P(T_f,y) = \left(K - f(T_f, y, Q)\right)^+,
\end{equation}
\noindent should be used, similar to how this is explained in \cref{asianE}.. Accordingly, this results only in a replacement
$ x_B(0) f^+(0) \to x_B(0) f^+(0) - x \left(K - f(0, x, Q)\right)^+$, while the derived equations remain the same. Once the price of the European option has been determined, the corresponding price of the American option can be found as described in \cref{amer3Mdes}.

\subsection{An explicit representation of $f(\tau, x_B(\tau), Q)$} \label{solForward}

In the continuation region, under a risk-neutral measure $\mathbb{Q}$ by a standard argument, \cite{andersen2010interest}, the ZCB price $F(t, y, Q )$  solves a linear PDE
\begin{equation} \label{PDE-42}
\fp{F}{t} - \alpha(t) y \fp{F}{y} + \frac{1}{2} \sigma^2(t) y^{2 \beta+2} \sop{F}{y} - [\barR^*(t) + y] F= 0,
\end{equation}
\noindent subject to the terminal
\begin{equation} \label{termZCB-42}
	F(Q,y,Q)  = 1,
\end{equation}
\noindent and the boundary conditions
\begin{equation} \label{bc1-42}
	F(t,y_l(t),Q)  = 1, \qquad  F(t, y, Q)\Big|_{y \to \infty} = 0.
\end{equation}

Again, we can apply Proposition ~\ref{prop1} to this problem to transform it into the following one
\begin{align} \label{wZCB}
\fp{U}{\tau} &= \frac{1}{2} \sop{U}{x} + \frac{b}{x} \fp{U}{x} + \lambda(\tau,x) - g(\tau,x) \left[ U + \frac{x_l(\tau)}{x} \right], \\
\lambda(\tau,x) &= \frac{x_l(\tau)}{x} \frac{1 - b}{x^2} + \frac{x'_l(\tau)}{x}, \nonumber \\
U(0,x) &= 1 - \frac{x_l(0)}{x}, \qquad U(\tau, x \uparrow \infty) = U(\tau, x_l(\tau)) = 0, \qquad
U(\tau,x) = u(\tau, x) - \frac{x_l(\tau)}{x}, \nonumber
\end{align}
\noindent and $y(\tau,x)$ is defined in \eqref{BessSource}.

The solution to this problem looks similar to that in \eqref{uBessFull}
\begin{align}  \label{uFfull}
u(&\tau, x) =  1 - \frac{x_l(\tau) }{x} + \int_0^\tau \int_{x_l(s)}^\infty \left[ \lambda(s, \xi) - g(s,\xi) u(s,\xi) \right] \varTheta_{|\nu|}(\tau - s, x, \xi, x_l(\tau) ) d\xi ds \\
&+ x^{-\nu} \Bigg\{ \frac{1}{2}\int_{0}^\tau x_l^{\nu + 1}(s) \Psi(s, x_l(s)) \varTheta_{|\nu|}(\tau - s, x, x_l(s), x_l(\tau) ) ds - K x_l(0) \int_{x_l(0)}^\infty \xi^{\nu} \varTheta_{|\nu|}(\tau, x, \xi, x_l(\tau) ) d\xi \Bigg\}. \nonumber
\end{align}

Differentiating both parts of this equation by $x$ and substituting $x = x_l(\tau)$ we obtain a LIVESK for the gradient $\Psi(\tau, x_l(\tau))$
\begin{align}  \label{uFPsi}
\Psi(\tau, x_l(\tau)) &=  (x_l(\tau))^{-\nu} \Bigg\{ \frac{1}{2}\int_{0}^\tau x_l^{\nu + 1}(s) \Psi(s, x_l(s)) \varTheta'_{|\nu|}(\tau - s, x_l(\tau), x_l(s), x_l(\tau) ) ds \\
&- K x_l(0) \int_{x_l(0)}^\infty \xi^{\nu} \varTheta'_{|\nu|}(\tau, x_l(\tau), \xi, x_l(\tau) ) d\xi \Bigg\} + \frac{1}{x_l(\tau) } \nonumber \\
&+ \int_0^\tau \int_{x_l(s)}^\infty \left[ \lambda(s, \xi) - g(s,\xi) u(s,\xi) \right] \varTheta'_{|\nu|}(\tau - s, x_l(\tau), \xi, x_l(\tau) ) d\xi ds. \nonumber
\end{align}

\cref{uFfull,uFPsi} is a system of two LIVESK for the unknown dependent variables $u(\tau, x), \Psi(\tau, x_l(\tau))$. It can be solved numerically, similar to how this is described in \cref{secSolAsian}. Finally, following the definition in \eqref{forwardDef}, this solution for the ZCB price (expressed in the variable $u(\tau,x)$) can be substituted into \eqref{forwardDef} to compute the forward price. Thus, this solution builds a map between the variables $(\tau, x)$ and $f(\tau,x,Q)$. A reasonable choice would be to set $Q > t_0$. Then, when solving \eqref{wEqF}, every time the boundary $x_B(\tau)$ is given, we can use this solution to find the corresponding $f(\tau, x_B(\tau), Q)$ and substitute it into the boundary conditions in \eqref{wEqF}.

As follows from the analysis of \cref{betam1}, for the case $\beta = -1$, which corresponds to the Hull-White model, the solution of \cref{uFfull,uFPsi} can be significantly simplified. Due to the properties of the Weber-Orr Theta function given in Proposition~\ref{prop2}, \eqref{uFPsi} becomes an explicit semi-analytical expression for $\Psi(\tau, x_l(\tau))$ since the terms in the RHS corresponding to $s \to \tau$ vanish while the terms with $s < \tau$ are already known at the time $\tau$. Similarly, \eqref{uFfull} in this case becomes an explicit semi-analytical solution for $u(\tau,x)$, for more details, see the end of \cref{betam1}.

\section{Combining two regions together}

As we have already mentioned, CME trades options on 3M and 1M SOFR futures. These options are linked to the contract month of the
future, i.e., to the beginning of the reference period (the reference quarter for options on 3M SOFR futures). The key difference to the specifications of options on 3M SOFR futures from that on 1M SOFR futures is that both the 1M SOFR futures contract and the options on it end trading on the same day, i.e., on the last business day of the contract month, while trading in options on 3M SOFR futures ends before the reference quarter begins. This difference in specifications has two major consequences, \cite{HugginsSchaller}:
\begin{enumerate}
\item Unlike options on 3M SOFR futures, options on 1M SOFR futures undergo the transmogrification into an exotic Asian option during their lives.

\item Unlike options on 3M SOFR futures, options on 1M SOFR futures provide a complete series of hedging instruments for
caps and floors.
\end{enumerate}
Also, the strike prices of options on 1M SOFR futures do not cover a very large range. and are close to ATM.

Below, we consider a pricing problem of these options in more detail using the results, obtained in the previous sections.

\subsection{Options on 1M SOFR futures} \label{1mAmerican}

\begin{figure}[!htbp]
\centering
\fbox{\includegraphics[width=0.6\textwidth]{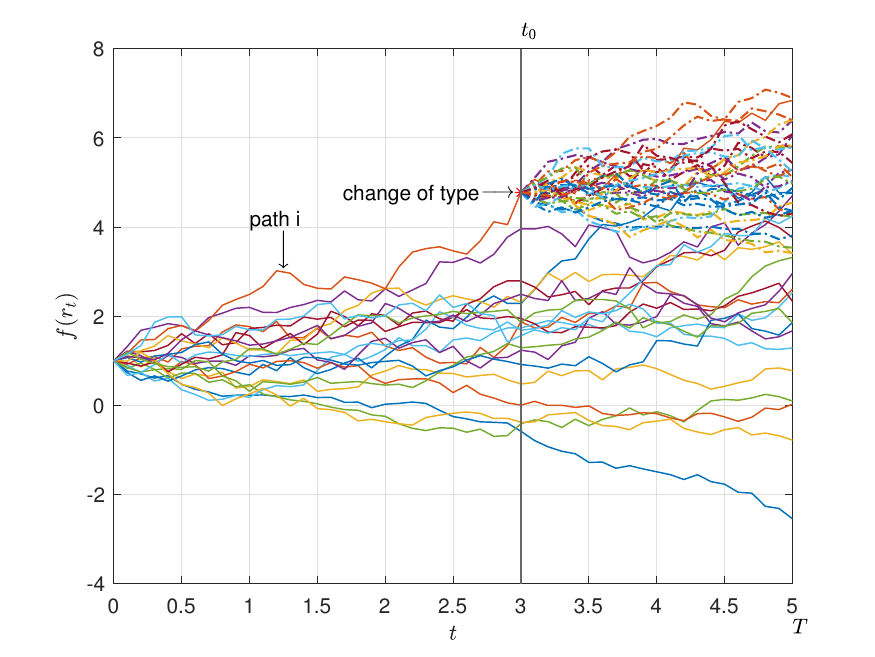}}
\caption{A schematic Monte Carlo approach to illustrate the pricing of options on 1M SOFR futures (here $f(r_t)$ is some function of $r_t$).}
\label{path}
\end{figure}

Since the 1M options on SOFR futures mandatory change their type during the option life, let us consider this process in more detail. Suppose that for pricing these options we use a backward (American) Monte Carlo simulation, \cite{Jackel2002}, and move along the $i$-th path representing a certain realization of the underlying process, see Fig.~\ref{path}. The paths are generated for two areas: $t < t_0$  where the underlying is the forward price of the 1M SOFR futures, and the reference period $t \ge t_0$ where the underlying is the daily SOFR. At the point $t=t_0$ the option changes its type from being an American one at $t < t_0$ to becoming an Asian-American option at $t_0 \le t \le T$.

Therefore, when running the backward step of the American Monte Carlo method, at the path $i$ and at the point $t=t_0$ there is a choice of how to proceed. The holder can exercise the American option a $t=t_0$ to get payoff $K - f^{1m}(0;t_0,T,r)$, where $f^{1m}(0;t_0,T,r)$ is the time $t_0$ rate of the 1M future with settlement at the time $T$, which depends on $r$ - the SOFR rate at $t_0$ along the $i$-th path. Alternatively, she can continue.

If we know the $t_0$-price $P(t_0, y(r), z(r))$ of the Asian-American Put option with maturity $T$ conditional on $r$ for the reference period, then the realized payoff of the option at $t=t_0$ is (see \cref{contCompound,singul})
\begin{equation} \label{combPayoff}
\Phi = \max(K - f^{1m}(0;t_0,T,r), P(t_0, y(r), z(r)), \qquad y(r) = r - \barR^*(t_0), \, z(r) = r.
\end{equation}

\subsubsection{Computing $\bm{P(t_0, y(r), z(r))}$ } \label{AmerSol}

To recap, in \cref{Asian} we have developed an approach which provides a semi-analytical solution for the exercise boundary $z_B(\tau, x)$ and the European option price $u(\tau,x,z)$ in both continuation and the entire regions. Once this is done, the solution to the American problem can also be found by using these equations since the solution is obtained recurrently in time.\footnote{Rigorously speaking, this converts the American option into the Bermudan one. However, the option price should converge to the corresponding American option value when the integration step $\Delta \tau \to 0$.} Namely, the integration \eqref{ODE} in $z$ can again start from the exercise boundary (instead of infinity as for the European options) since it is already known as well as the option value and the option Delta at this boundary. Since we integrate \cref{uBessFullNew2,uBessFullPsiRec} sequentially in time, at every step the computed value of $u(\tau,x,z_j)$ can be compared with the intrinsic value $K - z_B(\tau,x)$. This means that for pricing American options, instead of $\bar{u}(\tau,x,z_j)$ in \cref{uBessFullNew2,uBessFullPsiRec} one has to use
\begin{equation}
\bar{u} = \max(\bar{u}(\tau,x,z_j),K - z_j), \qquad \bar{u}_z = \max(\bar{u}_z(\tau,x,z_j),-1).
\end{equation}
This method is similar to pricing American options using a binomial tree method, or a simple version of the FD method (rather than a more sophisticated and accurate approach of \cite{Halluin2004}).

\subsubsection{The Put option price at $\bm{t=0}$}

Since this payoff in \eqref{combPayoff} is conditional on the value $r$ (or $y$) at the time $t = t_0$, the {\it European} option price $P_E$ at $t=0$ is given by the expectation of it over all possible realizations of the underlying process, i.e.,
\begin{align}
P_E = \EQ \left[ e^{- \int_0^{t_0} r(s) ds} \max \left(K - f^{1m}(0;t_0,T,r), P(T, y(r), z(r) \right) \, \Big| \, r_0 \right] ,
\end{align}
\noindent or, more explicitly
\begin{equation} \label{combInt}
P_E = \int_{r_l}^\infty e^{- \int_0^{t_0} r(s) ds} \max \left(K - f^{1m}(0;t_0,T,r), P(T, y(\chi), z(\chi) \right)
\Psi(0, r_0 | t_0, \chi) d\chi, \qquad r_l = \barR^*(0) + y_l(0),
\end{equation}
\noindent where $\Psi(0, r_0 | t_0, \chi)$ is the density of the underlying process. It turns out that all components of this formula under the integral can be found by using the approach described in \cref{Asian,American}. Indeed, the Asian-American option price can be found by using the method described in \cref{Asian} and also above in this Section. The density function at the interval $t \in [0,t_0]$ can be computed by using the approach described in \cref{American} if one uses the Dirac Delta function as the terminal condition and chooses $T_f = t_0$. Alternatively, one can solve the problem in \cref{American} and use the payoff in \eqref{combPayoff} instead of \eqref{tchw}. In the former case, an additional integral in \eqref{combInt} should be computed, while the problem of finding the density function is simpler due to the terminal condition being a Delta function. In the latter case, no additional integration is required. In both cases we set $Q = T$.

Now, for the American option at the interval $t \in [0, t_0]$, we need to use the second approach. To recap, again this approach is valid only in the continuation region. However, it can be extended to provide an European Put option price in the whole domain of $z$, as this was described at the end of \cref{Asian}. Once this is done, the solution to the American problem can also be found as described in \cref{AmerSol}.

As mentioned in \cite{Skov2021}, the settlement of the 1M federal funds futures is based on the same specifications and the pricing formula is, therefore, valid for both SOFR and federal funds 1M futures traded at the CME.

A more detailed analysis of the results obtained by using this formalism will be presented elsewhere.

\subsection{Options on 3M SOFR futures} \label{amer3Mdes}

The 3M futures contract is based on the daily compounded reference rate during the contract quarter
\begin{equation}
R^{3 m}(t_0, T) = \frac{1}{T-t_0} \left( \prod_{i=1}^N \left[ 1+d_i R_{d_i}\left(t_i\right)\right] - 1 \right), \qquad
i \in 1,\ldots, N, \quad t_0 \leq t_1,\ldots,t_N \leq T,
\end{equation}
\noindent where $R_{d_i}(t_i)$ denotes the realized overnight rates in the reference quarter, and $d_{t_i}$ is the amount of days to which $R_{d_i}(t_i)$ applies. It can be approximated by the continuously compounded rate as, \cite{Mercurio2018}
\begin{equation}
R^{3 m}(t_0, T) \approx \frac{1}{T-t_0}\left(e^{\int_{t_0}^T r_s d s}-1\right)
\end{equation}
As shown in \cite{Skov2021}, the accuracy of the continuous approximation is of the order of $10^{-7}$ for all open contracts. The time $t$ rate of the 3M future starting to accrue at time $t_0$ and with settlement on time $T$ (so $T-t_0 = 0.25$) is given by
\begin{equation}
f^{3 m}(t ; t_0, T) = \frac{1}{T-t_0}\left(\mathbb{E}^Q\left[e^{\int_{t_0}^T r_s d s} \mid \mathcal{F}_t\right]-1\right)
\end{equation}

Note, that if $t_0 < t$ and part of the underlying rate has already accrued, this can be accounted for by using the discrete compounding, \cite{Skov2021}
\begin{align}
f^{3 m}(t;t_0,T) &= \frac{1}{T-t_0}\left\{ \left(\prod_{i=1}^{N_0}\left[1+d_i R_{d_i}\left(t_i\right)\right]\right) \mathbb{E}^Q \left[e^{\int_t^T r_s d s} \mid \mathcal{F}_t\right] - 1 \right\}, \\
i & \in 1,\ldots,N_0, \quad t_0 \leq t_1,\ldots,t_{N_0} \leq t, \nonumber
\end{align}
\noindent where $R_{d_i}(t_i)$ denotes the $\mathcal{F}_t$-measurable realized overnight rates.

The price of an American option written on $f^{3 m}(t;t_0,T)$ with $t < t_0$ can be computed by using the approach of \cref{American} with the extension described in \cref{AmerSol}. Again, a more detailed analysis is left to be presented elsewhere.

\section{Numerical example} \label{experiments}

In this section, our approach is illustrated by a numerical example where, for simplicity, we set $\beta = -1$, hence reducing the full CEV model to the time-dependent Hull-White model. We also solve only the problem described in \cref{Asian}, assuming that the initial parameters $y, \barr(t_0), R^*(t_0)$ are given.

It could be observed, that almost all the existing literature on pricing arithmetic Asian options of the American exercise style with fixed strike uses the Black-Scholes model with fixed coefficients for the underlying stock process. Thus, to compare our results with those in, e.g., \cite{ZvanForsythVetzal1997}, we mimic their settings by choosing $\alpha(t) = \alpha, \, \sigma(t) = \sigma$, i.e., by setting parameters of the model as in Table~\ref{tab1}.

\begin{table}[!htb]
\begin{center}
\begin{tabular}{|c|c|c|c|c|c|c|c|}
\hline
$\alpha$ & $\sigma$ & $\barR(t)$ & $T$ & $y$ & $x_{max}$   \\
\hline
-0.1 & 0.2 y  & -0.01 & 0.25 & 100 & 1000 \\
\hline
\end{tabular}
\caption{Parameters of the test.}
\label{tab1}
\end{center}
\end{table}
Here, $\sigma(t)$ is the normal volatility, therefore, we multiply it by the initial spot level $y$ to approximately mimic the corresponding log-normal volatility. Also, we run the test for a set of strikes $K \in [90, 95, 100, 105, 110, 120]$. Since variables $y$ and $x$ are defined at a semi-infinite interval, when doing numerical computations, we truncate the upper limit and set it to the value $x_{max}$.

We compute the exercise boundaries for the Asian-American Put option by using the numerical scheme in \cref{finalization}. The corresponding integrals in the time $\tau \in [0,\tau_m]$ are approximated by using the Simpson quadratures on a 3D grid, which contains $N_t = 40$ nodes in $\mathbb{T}$. The integrals in $x$ and $z$ are computed by using the trapezoid rule. We take $N_x = 200$ nodes in $\mathbb{X}$, as it turns out that the further increase of $N_x$ practically doesn't affect the results. We constructed unified grids for $\mathbb{T}$ and $\mathbb{Z}$ (with $N_z = 10$), while for $\mathbb{X}$ a non-uniform gird is constructed with nodes compressed close to $y$, less compressed close to $x_l(\tau)$, and rarefied as the running $x$ tends to $x_{max}$, see, e.g., \cite{ItkinBook} among others. Thus, the accuracy of the scheme is $O(1/N^4_t)$ in time for the temporal integrals from $0$ to $\tau_m$, $O(\left[\frac{\tau - \tau_m}{T}\right]^2)$ for the last temporal interval, and $O(1/N_z)$ in $z$\footnote{This accuracy could be easily increased, if necessary, by using the one-sided finite difference of the second order when approximating the first derivative in $z$, \cite{ItkinBook}. }.

The results are obtained in Matlab using two Intel Quad-Core i7-4790 CPUs, each 3.80 Ghz. Despite the fact that our Matlab code can be naturally vectorized, we didn't do it since in this example we used a simple method that can be approved in many different ways, i.e., see \cite{ItkinMuravey2024jd} and reference therein, among others.

Since $\beta = -1$, computation of the Weber-Orr Theta function simplifies as this is described in Proposition~\ref{prop2}. The workflow of computations is presented in Fig.~\ref{Algo}.
\begin{figure}[!htb]
\begin{center}
\begin{minipage}{0.9\textwidth}
\begin{algorithm}[H]
 \hspace*{\algorithmicindent} \textbf{Input:} Data from Table~\ref{tab1} \\
 \hspace*{\algorithmicindent} \textbf{Output:} $z_B(\tau, x), u(\tau,x,z)$
\begin{algorithmic}[1]
    \Procedure{Asian-American Pricing}{} \CommentF{As per \cref{Asian}}
       \State \emph{{\bf Initialization}: All values of $\Psi(\tau, x_l(\tau), z)$, $z_B(\tau, x)$ and $u(\tau, x,z)$ at $\tau = 0$.}
        \For { $s \in [0, \tau]$ }                  \CommentF{Main loop in $\tau$}
            \For { $x \in [x_l(s), x_{max}]$} \CommentF{Run the first internal loop in $x$}
                \State 1. Compute $z_B(\tau,x)$ from \eqref{zbFinal}
                \State 2. Compute $\Psi(\tau, x_l(\tau), z_B(\tau, x))$ in \eqref{uBessFullPsiBrec}
             \EndFor

             \For {$\xi \in [z_B(s, x), z]$} \CommentF{Run the second internal loop in $x$ and $z$}
                    \State Find $\Psi(\tau, x_l(\tau), \xi)$ from \eqref{uBessFullPsiRec}
             \EndFor

             \For {$x \in [x_l(s), x_{max}]$} \CommentF{Run the third internal loop in $x$ and $z$}
                \For {$\xi \in [z_B(s, x), z]$}
                    \State Find $\bar{u}(\tau, x, \xi_i)$ from \eqref{uBessFullNew2}
                \EndFor
             \EndFor
        \EndFor
    \EndProcedure
\end{algorithmic}
\caption{Pricing algorithm.}
\end{algorithm}
\end{minipage}%
\end{center}
 \caption{Algorithm of pricing arithmetic Asian options of the American exercise style with fixed strike.}
 \label{Algo}
\end{figure}

There are few comments about the initialization of this computational scheme. As shown in Fig.~\ref{Algo}, the initialization step requires providing all values for $\Psi(\tau, x_l(\tau), z), \, z_B(\tau, x)$ and $u(\tau, x, z)$ at $\tau = 0$. For $z_B(0,x)$ this value is given by \eqref{tc0}, so $z_B(0,x) = K, \, \forall x \in [x_l(0), \infty)$, where $t = t(\tau), \, t(0) = T$. This also means that $P(t(0), y, z_B(t(0), y)) ) = 0$. Next, by definition in \cref{lPsi,bc1}, we have
\begin{align}
\Psi(0, x_l(0), z) &= \left. \fp{u(0, x, z)}{x} \right|_{x = x_l(0)} = \left. y_l^{\beta+1}(T) \fp{}{y}P(T, y, z) \right|_{y = y_l(T)}, \\
P(T, y_l(T), z) &= K. \nonumber
\end{align}
\noindent since $\phi(T) = 0, \, F(0) = 1$. Substituting $\tau \to 0$ into \eqref{uBessFullPsiBrec}, we obtain
$\Psi(0, x_l(0), z_B(0,x)) = -K/x_l(0)$. To recall, this is not the gradient at the exercise boundary, but rather the gradient at the left boundary in variable $y$, which, by its financial nature, should be sufficiently smooth.

Accordingly, based on \eqref{uBessFullNew}, we have $\bar{u}(0, x, z) = 0$, i.e., $u(0,x,z) = K - z_B(0,x) = 0$, which agrees with \eqref{tc0} via the definition of $u(\tau, x, z)$ in \cref{tr1,tr2}.

\begin{figure}[!htp]
\centering
\fbox{\includegraphics[width=0.7\textwidth]{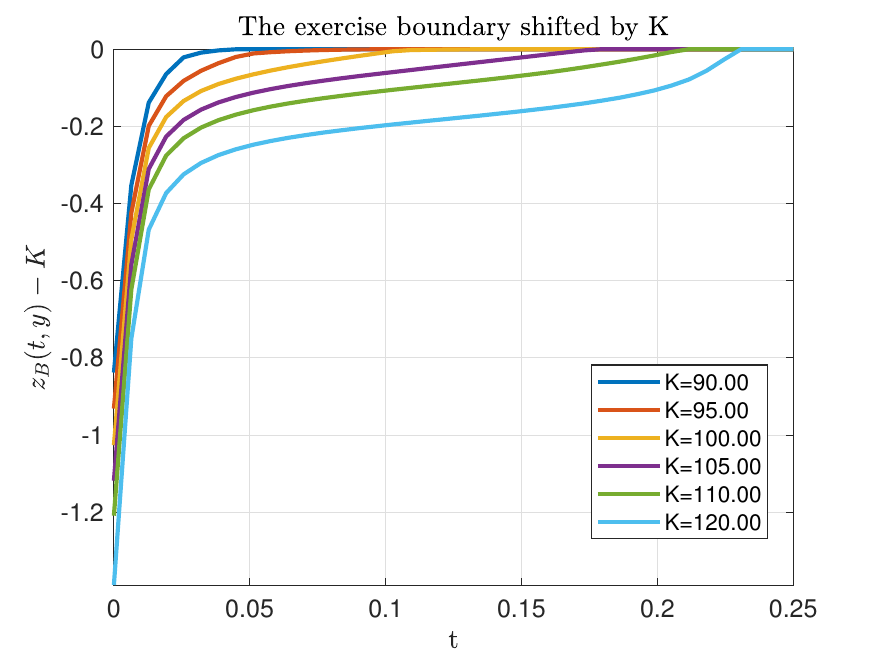}}
\caption{The exercise boundary $z_B(t, y)$ computed for an Asian-American Put option.}
\label{zB1}
\end{figure}

\paragraph{Results.} In Fig.~\ref{zB1} the exercise boundary $z_B(t, y)$ computed in this test is presented as a function of the time $t$. The elapsed time per strike to compute the whole 3D exercise boundary (i.e., using every of the $N_x$ nodes as the initial value of  $x$) is 2 seconds. As expected, the exercise boundary demonstrates a jump at $t \to t_0$ and tends to $z_B(t_0, y) = K$. This is because of the jump in the definition of $C(\tau,x,z)$ in \eqref{BessSource} at $t = t_0$ which follows from the standard assumption in \eqref{singul} taken to resolve the singularity in the PDE in \eqref{BessSource}.

\begin{figure}[!htp]
\centering
\fbox{\includegraphics[width=0.7\textwidth]{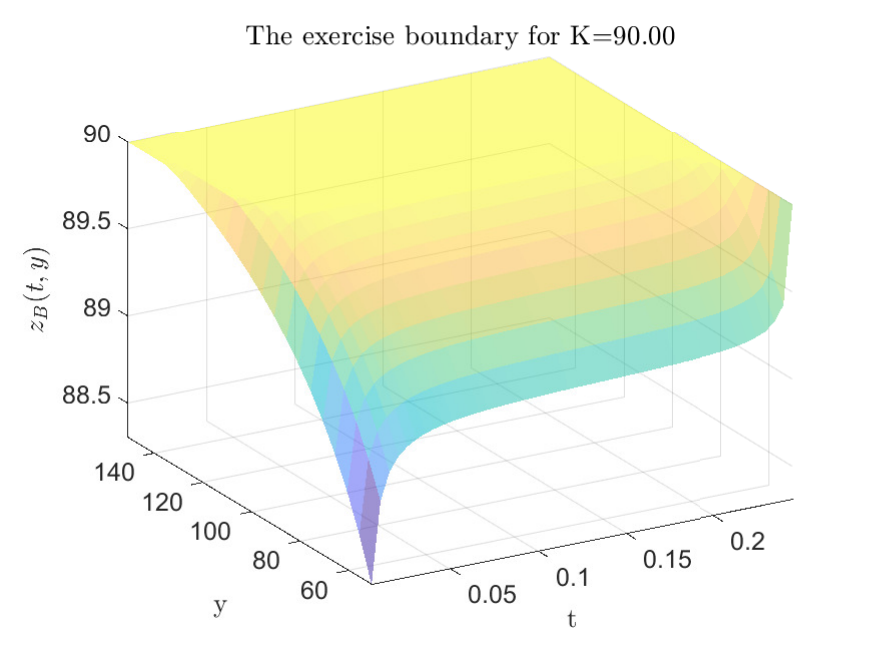}}
\caption{The 3D exercise boundary $z_B(t, y)$ computed for an Asian-American Put option with $y \in [50,150]$ and $K = 90$.}
\label{zB3D}
\end{figure}

\begin{figure}[!htp]
\centering
\fbox{\includegraphics[width=0.7\textwidth]{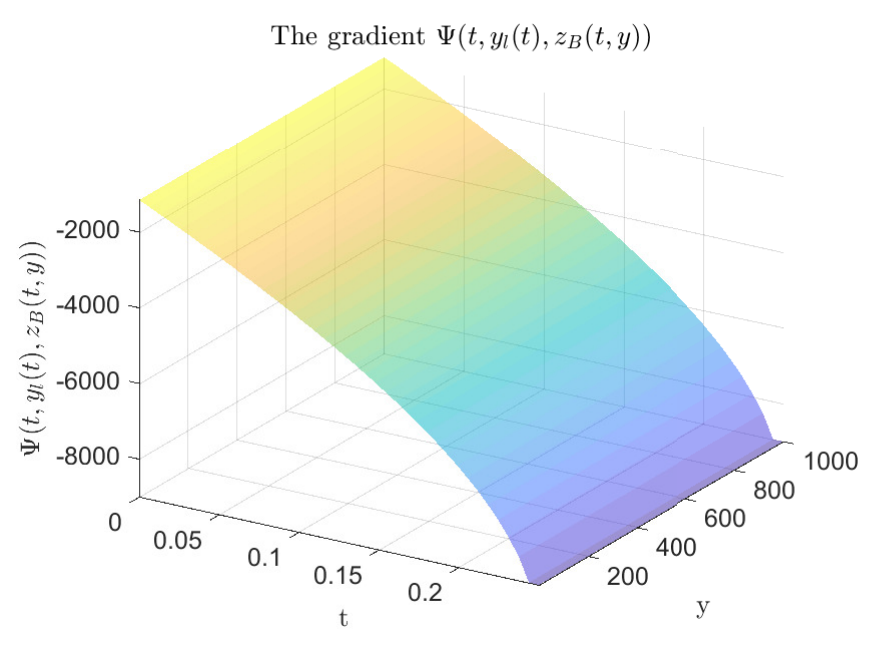}}
\caption{The 3D plot of $\Psi(t, y_l(t), z_B(t, y))$ computed for an Asian-American Put option with $y \in [50,150]$ and $K = 90$.}
\label{PsiZB}
\end{figure}

In Figs.~\ref{zB3D},\ref{PsiZB}, the same results are shown as a 3D plot of $z_B(t,y)$ at $K = 90$ and another 3D plot of $\Psi(t, y_l(t), z_B(t, y))$ both computed for an Asian-American Put option with $y \in [50,150]$.

\begin{figure}[!htb]
\begin{center}
\hspace*{-0.3in}
\subfloat[]{\includegraphics[width=0.57\textwidth]{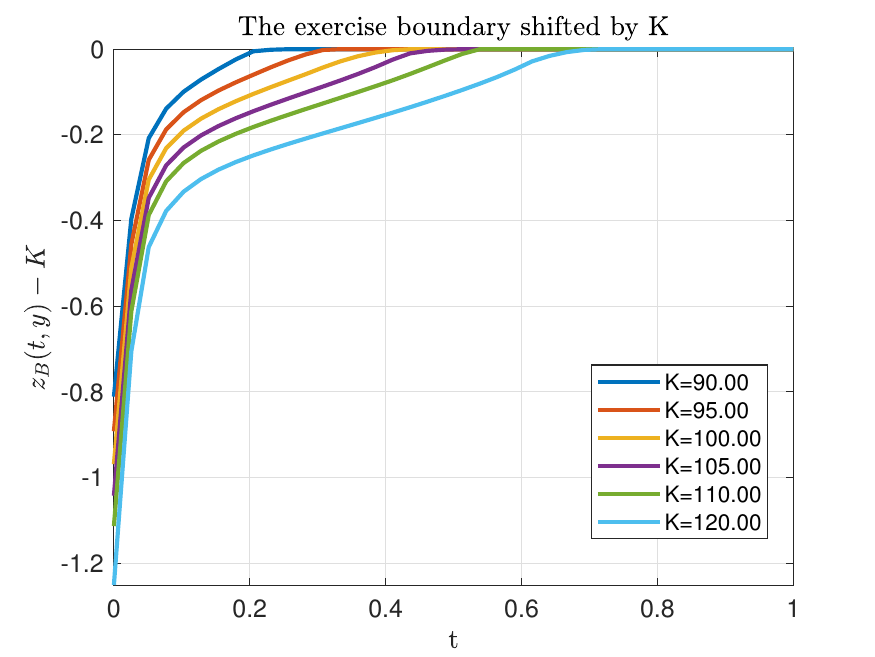}}
\hspace*{-0.3in}
\subfloat[]{\includegraphics[width=0.57\textwidth]{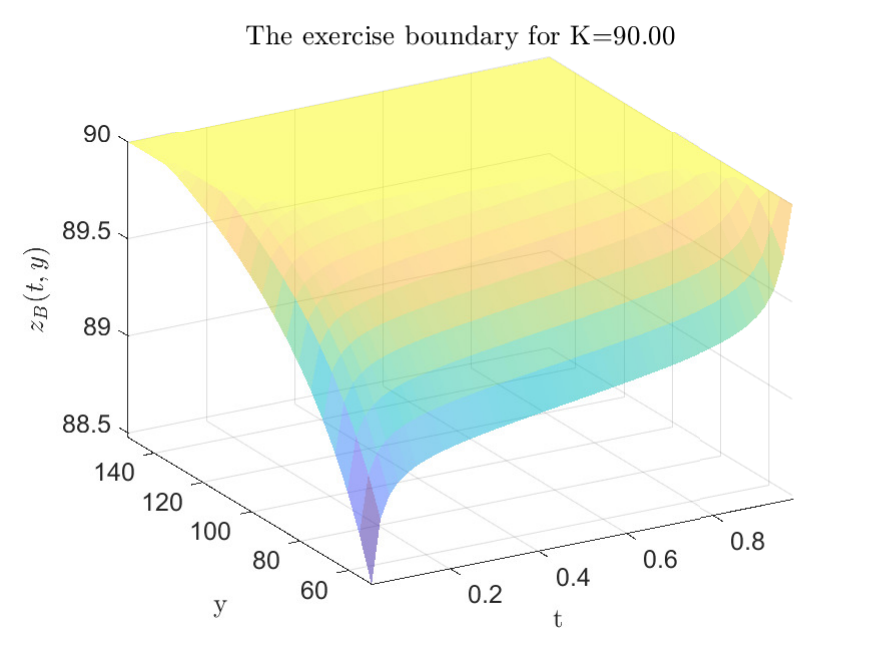}}
\end{center}
\caption{The exercise boundary for an Asian-American Put option in the test with $T=1$ year: a) the exercise boundary shifted by $K$; b) the 3D exercise boundary $z_B(t, y)$ computed with $y \in [50,150]$ and $K = 90$.}
\label{Fig1year}
\end{figure}

\begin{figure}[!htb]
\begin{center}
\hspace*{-0.3in}
\subfloat[]{\includegraphics[width=0.57\textwidth]{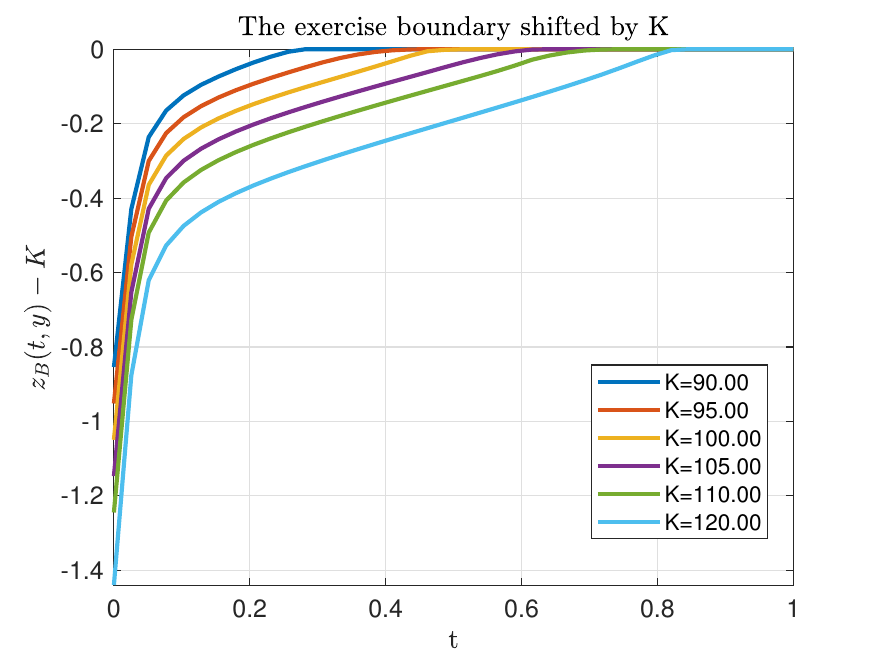}}
\hspace*{-0.3in}
\subfloat[]{\includegraphics[width=0.57\textwidth]{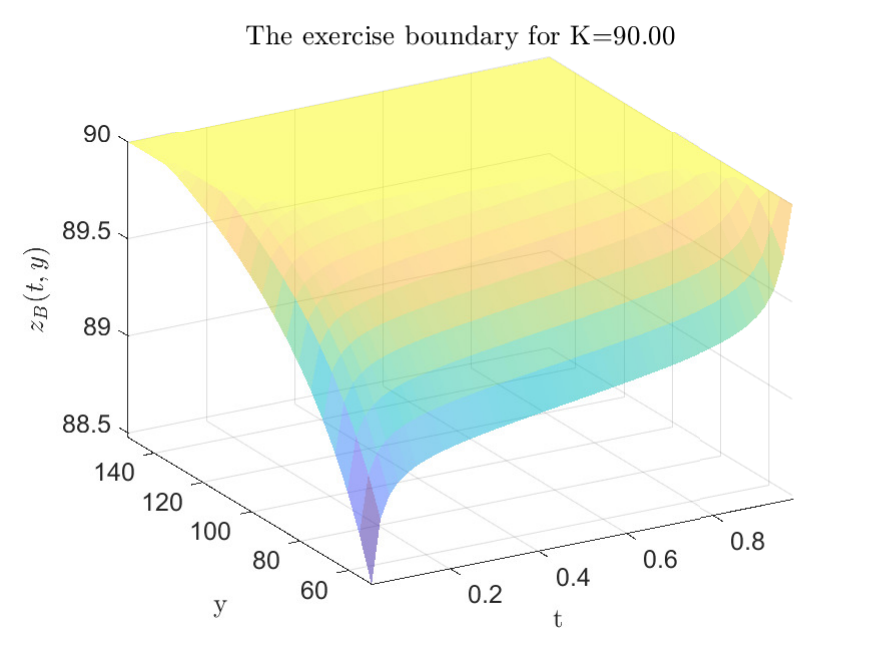}}
\end{center}
\caption{The exercise boundary for an Asian-American Put option in the test with $T=1$ year and no $y$ in the killing term in \eqref{PDE}: a) the exercise boundary shifted by $K$; b) the 3D exercise boundary $z_B(t, y)$ computed with $y \in [50,150]$ and $K = 90$.}
\label{Fig1yearEq}
\end{figure}

Similar results, but for $T=1$ year, are presented in Fig.~\ref{Fig1year}. On the other hand, the PDE in \eqref{PDE} differs from that used for equity options since the last killing term is proportional to $y$ while for equities this should be an interest rate independent of $y$. To see the influence, we again run the same test, but set $y=0$ in the killing term in \eqref{PDE}. The results are presented in Fig~\ref{Fig1yearEq}.

\begin{figure}[!htb]
\begin{center}
\hspace*{-0.3in}
\subfloat[]{\includegraphics[width=0.57\textwidth]{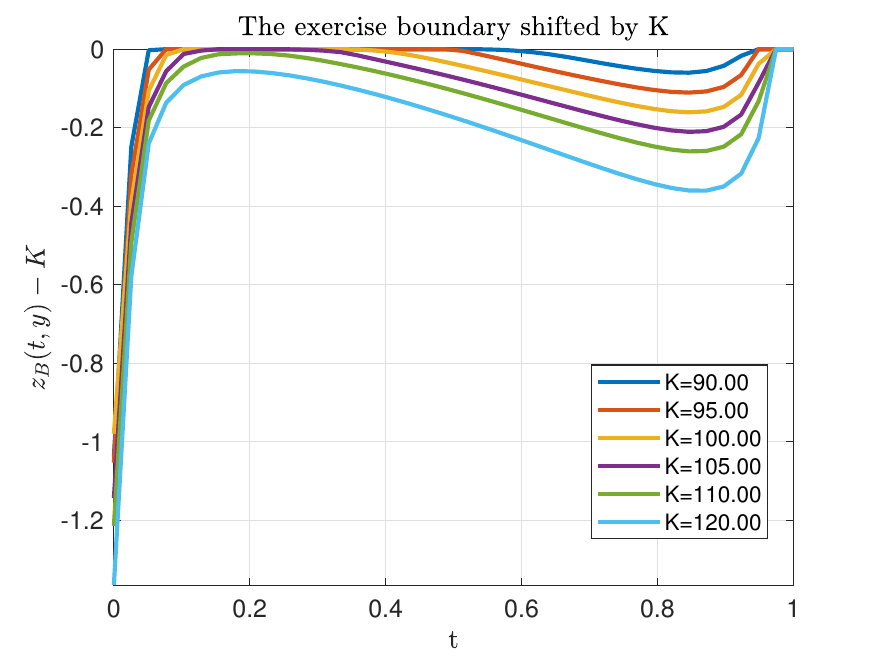}}
\hspace*{-0.3in}
\subfloat[]{\includegraphics[width=0.57\textwidth]{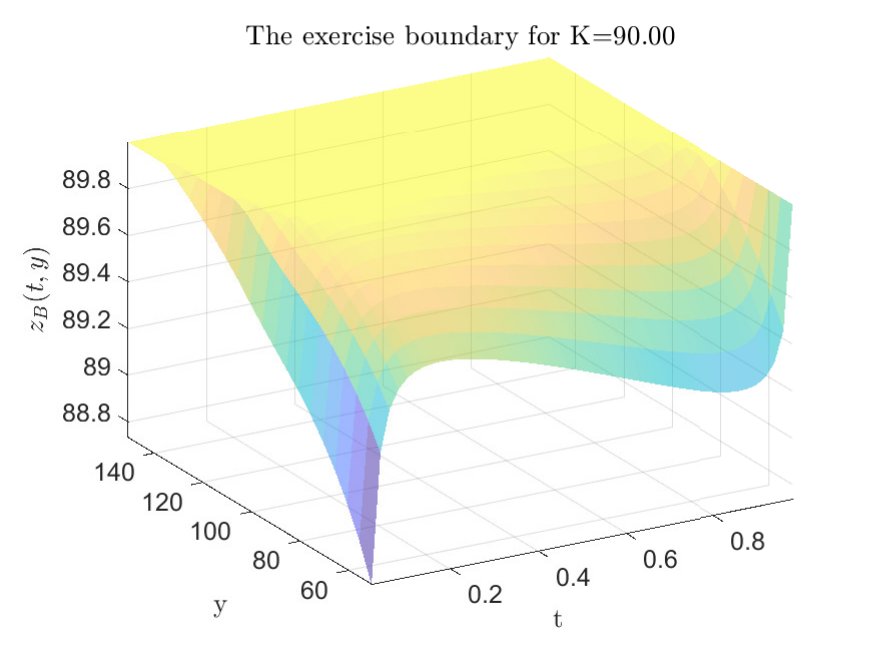}}
\end{center}
\caption{The exercise boundary for an Asian-American Put option in the original model with mean reversion $\alpha = 0.1$, and $T=1$ year: a) the exercise boundary shifted by $K$; b) the 3D exercise boundary $z_B(t, y)$ computed with $y \in [50,150]$ and $K = 90$.}
\label{figOrig}
\end{figure}

\begin{figure}[!htp]
\centering
\fbox{\includegraphics[width=0.7\textwidth]{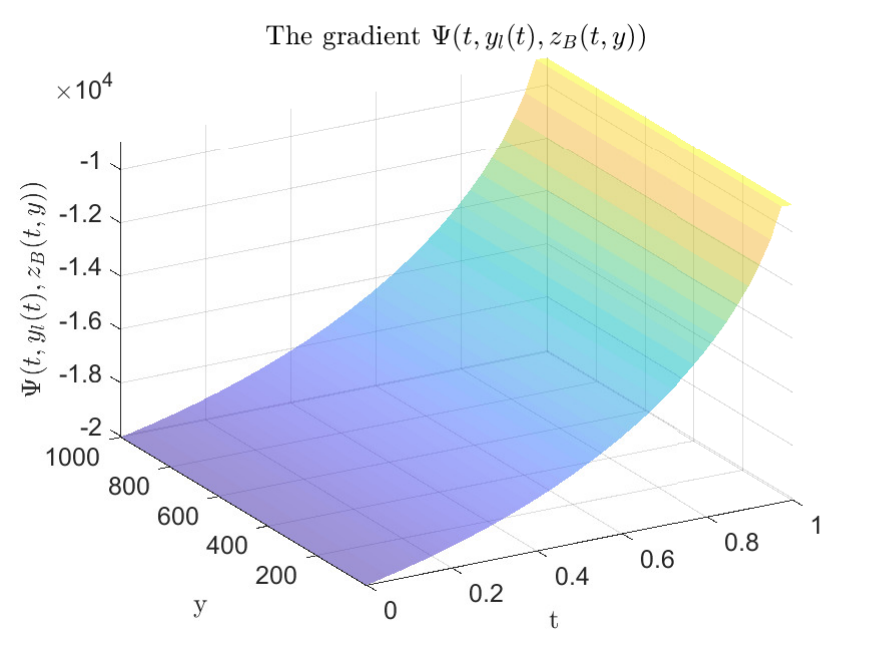}}
\caption{The 3D plot of $\Psi(t, y_l(t), z_B(t, y))$ computed for an Asian-American Put option with $y \in [50,150]$ and $K = 90$ in the original model with $T$ = 1 year.}
\label{figOrigPsi}
\end{figure}

Finally, we run this test for the original model with mean reversion, i.e., by setting $\alpha = 0.1$ and using the original killing term. The results are presented in Fig.~\ref{figOrig}. It can be seen that the exercise boundary has a local minima in $t$ which is not a typical behavior for American equity options (they don't have it, e.g., see plots in \cite{ItkinMuravey2024jd}).  If we set $y=0$ in the killing term in \eqref{PDE}, the well's depth decreases, but the shape of the exercise boundary remains unchanged. Also, the gradient function $\Psi(t, y_l(t), z_B(t, y))$ depicted in Fig~\ref{figOrigPsi} now increases its absolute value while decreasing in level.

Also, in our test the exercise boundary doesn't deviate much from the strike level. This can be explained by the fact that, despite using the Hull-White model with adjusted normal volatility to match the corresponding log-normal model, the volatility still doesn't depend on the underlying.

Once the $z_B(\tau,y)$ and $\Psi(t, y_l(t), z_B(t, y))$ have been found, the option value $u(\tau,y,z)$ can be computed by using \eqref{uBessFullNew2} and \eqref{uBessFullPsiRec}. To recap, the PDE in \eqref{PDE} is valid only in the continuation region. However, it can be adjusted to provide an European Put option price in the whole domain of $z$ as described at the end of \cref{Asian}. Once this is done, the solution to the American problem can also be found as described at the end of \cref{1mAmerican}.

\begin{figure}[!htp]
\centering
\fbox{\includegraphics[width=0.7\textwidth]{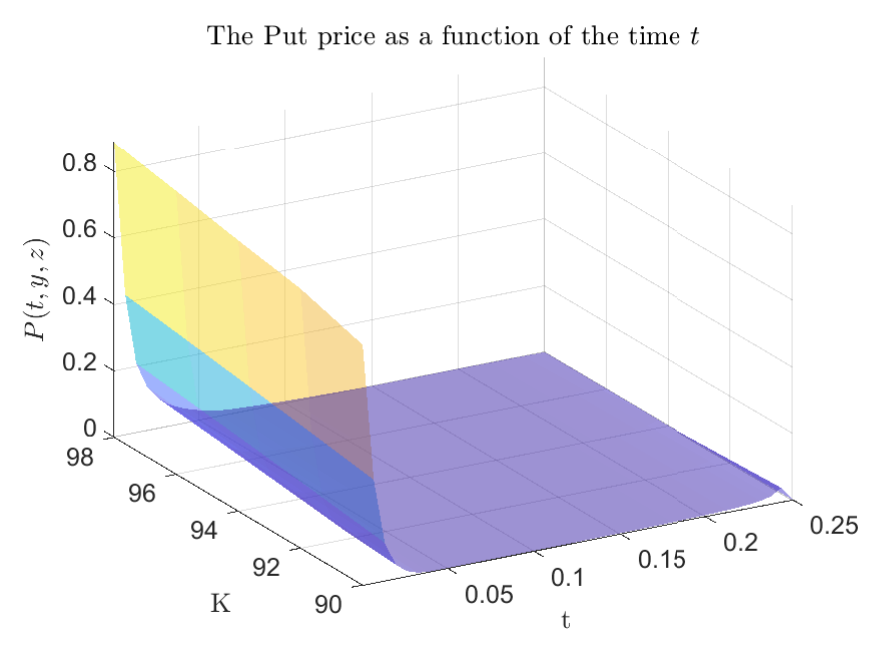}}
\caption{The Asian-American put option price as a function of the strike $K$ and the time $t$ for the original model.}
\label{putPrice}
\end{figure}

In Fig.~\ref{putPrice} the Asian-American Put option price computed by using this approach is presented as a function of $K$ and the time $t$, assuming $t_0 = 0$ and $\alpha = 0.1$. For the reasons mentioned at the end of \cref{Asian}, we consider only the continuation region. The typical elapsed time per strike is 19 seconds \footnote{Further optimization of performance by using Matlab vectorization makes it possible to reduce this time by approximately another 7 seconds. However, at the moment, we are not focused on this too much.}

\begin{figure}[!htp]
\centering
\fbox{\includegraphics[width=0.7\textwidth]{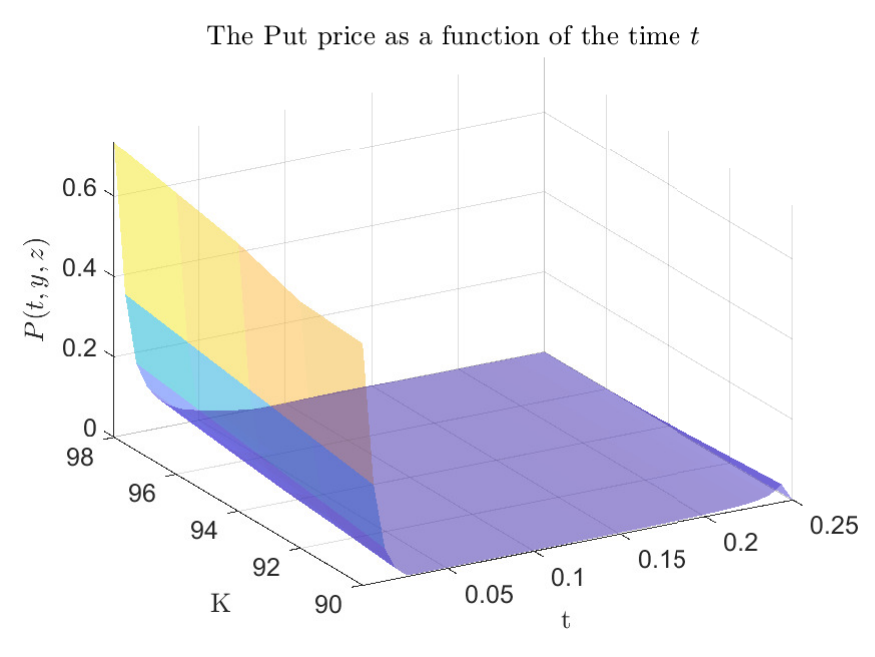}}
\caption{The Asian-American put option price as a function of the strike $K$ and the time $t$ for the "Equity-like" model with $y=0$ in the killing term of \eqref{PDE}.}
\label{putPriceEq}
\end{figure}

A similar plot is presented in Fig.~\ref{putPriceEq} where again we mimic the equity options by setting $y=0$ in the killing term and $\alpha = -0.1$ in \eqref{PDE}.  A comparison of these prices with those of Asian-American Put options computed by using a log-normal (Black-Scholes) model with the same model parameters and a CRR tree method is given in Table~\ref{compPrice}. Again, the option prices in the log-normal model grow faster than that in the normal model, even when an approximation of the log-normal volatility is used.

\begin{table}[!htb]
 \centering
 \begin{tabular}{|l|rrrrrr|}
 \toprule
\multicolumn{1}{|c|}{\multirow{2}[4]{*}{Put option}} & \multicolumn{6}{c|}{K} \\
\cmidrule{2-7}          & 90    & 92    & 94    & 96    & 98    & 100 \\
\hline
Asian-American  & 0.5575 & 0.5189 & 0.5897 & 0.6364 & 0.6635 & 0.6957 \\
European (Black-Scholes) & 0.0431 & 0.1176 & 0.2814 & 0.6014 & 1.1682 & 2.0883 \\
\bottomrule
 \end{tabular}%
\caption{Comparison of Asian-American Put option prices obtained by using our approach with those obtained by using a CRR tree method for the log-normal (Black-Scholes) model with the same model parameters.}
\label{compPrice}
\end{table}%

Note, that when computing an Asian-American option price with the same maturity by using the FD method of \cite{ZvanForsythVetzal1997} with $N_x = 40$ and $N_z$ = 45, the typical elapsed time is about 60 seconds for a single strike $K$. Therefore, even for the values of $N_t, N_x, N_z$ used in the above test (so, the total number of nodes $N_t N_x N_z$ in the 3D space  $[0,\tau(t_0)] \times [x_l(\tau),x_{\max}] \times [z_B(\tau,x),z]$  is almost equal for both methods) our method is faster.

However, it has to be emphasized that in a general case of an arbitrary value of $\beta$ the calculation of the Weber-Orr Theta function could be expensive. In our test example we used the particular value $\beta = -1$, which reduces this to computing two exponents. As a possible resolution, given $\beta$ the Weber-Orr Theta function can be computed offline and tabulated on a grid of input parameters. Then, the solution of the whole problem will have the same numerical complexity as in our test example.

\section{Conclusion} \label{Conclusion}

In this paper, we propose a semi-analytical approach to pricing options on SOFR futures. These options change their type at the beginning of the reference period $SR_{\mathrm{start}}$: Before this point in time, the option is settled as the American type written on a SOFR forward price as an underlying, and after this point, this is an arithmetic Asian option with an American style exercise written on the daily SOFR rates. Aside from problems with the existing numerical approaches for pricing options on SOFR futures, let us summarize what is achieved in this paper.

As mentioned in the Introduction, the GIT method for pricing American options has been previously proposed by one of the authors (in multiple papers co-authored with P.~Carr and with D.~Muravey, \cite{CarrItkin2020jd, ItkinMuravey2024jd, Itkin2024jd}). For various one-factor pure diffusion or even jump-diffusion models, a LIVESK for the option price was derived, supported by a nonlinear algebraic equation for the exercise boundary. This was an extension of the GIT method previously used for semi-analytical pricing of various barrier options, \cite{ItkinLiptonMuraveyBook}. However, this method was never applied to Asian options. Therefore, here, for the first time, we solve this problem by providing a semi-analytical solution for arithmetic Asian options with an American style exercise where the underlying follows the CEV process. The term "solve" means the following: i) we provide  a semi-analytical expression for the exercise boundary of the Asian-American Put option for any strike; ii) we also provide  a semi-analytical expression for the Asian-European and Asian-American Put option prices valid when $z \ge K$. However, we leave an extension of this approach to the case $z < K$ to be published elsewhere.

We also provide a solution to the other problem raised when pricing options on SOFR futures, which is pricing American options on SOFR forward prices. A similar problem with the underlying being a ZCB and for the Call option has already been solved in \cite{ItkinMuravey2024jd}. Here, we price an American Put option written on the SOFR forward rate by extending the approach of that paper.

The final solution to the problem is given in terms of the Weber-Orr Theta function, which was implicitly introduced in \cite{CarrItkinMuravey2020}, but is recognized as a separate special function with various extra properties just in this paper. It turns out, that for some particular values of $\beta$, e.g., $\beta = -1$ (which translates the CEV model into the Hull-White model), the representation of the Weber-Orr Theta function significantly simplifies since the Bessel functions could be expressed in terms of exponentials. Luckily, we prove that even in a general case with an arbitrary value of $\beta$, the aforementioned LIVESKs also translate into an explicit semi-analytical and recurrent solution to the problem. Hence, it doesn't require any numerical method except a pure computation of quadratures. Nevertheless, computation of the Weber-Orr Theta function per se could be tricky for extreme values of arguments because it requires computing  highly oscillating integrals and their derivatives. Fortunately, many of those problems can be automatically resolved with modern software, e.g., Wolfram Mathematica, but not all.

While all these results are new and important, they were obtained as auxiliary results to solve the problem of pricing options on SOFR futures. Our approach provides a pleasant alternative to pure numerical (FD and MC) methods, because we compute the exercise boundary explicitly. Therefore, jumps in some option Greeks at this boundary are naturally resolved within our method, while other methods compute it implicitly and, thus, introduce a bigger computational error. As mentioned, it is even possible to combine the FD and our methods, meaning one can first find the exercise boundary $z_B(\tau, x)$ by using our approach, and then in the continuation region solve \eqref{PDE} with the European payoff by using the FD method.

In Introduction, we cited \cite{HugginsSchaller}, who are concerned that  the transfer from LIBOR to SOFR has resulted in a situation in which the options of the key money market (i.e., futures on the reference rate) are options without any pricing model available. Therefore, the trading in options on 3M SOFR futures currently ends before their reference quarter starts, to eliminate the final metamorphosis into exotic options. We hope that our new approach can help with resolving this problem in an efficient manner.

\section*{Acknowledgments}

We are grateful to Dmitry Muravey for various fruitful discussions, and Colin Turfus, Alex Veygman and Thomas Schmelzer for useful comments.

\printbibliography[title={References}]

%

\end{document}